\documentclass[aps,prc,twoside,twocolumn,nofootinbib,10pt,showpacs,floatfix]{revtex4-1}
\usepackage{amsmath,amssymb}
\usepackage{graphicx,bm}
\usepackage{slashed}
\usepackage{epstopdf}
\usepackage[usenames]{color}
\usepackage{float}
\usepackage{hyperref}
\usepackage{subfigure}
\usepackage{rotating}
\usepackage{color}
\usepackage{multirow}
\usepackage{dcolumn}
\usepackage{overpic}
\usepackage{booktabs}
\usepackage{makecell}
\usepackage{diagbox}
\usepackage{array}
\newcolumntype{|}{!{\vline}}
\begin{document}
\title{Heavy flavor pentaquarks with four heavy quarks}
\author{Hong-Tao An$^{1,2}$}\email{anht14@lzu.edu.cn}
\author{Kan Chen$^{1,2,3,4}$}\email{chenk$_$10@pku.edu.cn}
\author{Zhan-Wei Liu$^{1,2,5}$}\email{liuzhanwei@lzu.edu.cn}
\author{Xiang Liu$^{1,2,5}$}\email{xiangliu@lzu.edu.cn}
\affiliation{
$^1$School of Physical Science and Technology, Lanzhou University, Lanzhou 730000, China\\
$^2$Research Center for Hadron and CSR Physics, Lanzhou University and Institute of Modern Physics of CAS, Lanzhou 730000, China\\
$^3$Center of High Energy Physics, Peking University, Beijing 100871, China\\
$^4$School of Physics and State Key Laboratory of Nuclear Physics and Technology, Peking University, Beijing 100871, China\\
$^5$Lanzhou Center for Theoretical Physics, Key Laboratory of Theoretical Physics of Gansu Province, and Frontiers Science Center for Rare Isotopes, Lanzhou University, Lanzhou 730000, China}

\date{\today}
\begin{abstract}
In this work, we carry out the study of heavy flavor pentatuarks with four heavy quarks, which have typical $QQQQ\bar q$ configuration. Within the Chromomagnetic Interaction model, the mass spectrum of these discussed $QQQQ\bar q$ pentaquarks is given. In addition to the mass spectrum analysis, we also illustrate
their two-body strong decay behavior by estimating some ratios of decay channels.
By these effort, we suggest that future experiment should pay attention to this kind of pentaquark.
\end{abstract}
\maketitle
\section{Introduction}\label{sec1}

On March 3 2021, LHC announced that 59 new hadrons were reported over the past decade \cite{LHCb59:2021}. The present situation of hadronic states is far beyond what Gell-Mann and Zweig thought possible \cite{GellMann:1964nj,Zweig:1981pd,Zweig:1964jf}. Among these observed states, these $P_c$ states existing in the $\Lambda_{b}{\to}{J/\psi}K^{-}p$ decay are good candidate of exotic molecular pentaquark \cite{Aaij:2015tga,Aaij:2016phn,Aaij:2019vzc,Chen:2015loa,He:2015cea,Chen:2015moa,Chen:2019asm,Wang:2019ato,He:2019rva,Liu:2019zvb,Burns:2019iih}. The name {\it pentaquark} was firstly proposed in Refs. \cite{Gignoux:1987cn,Lipkin:1987sk}, and there were many theoretical explorations of pentaquark \cite{Leandri:1989su,Genovese:1997tm,Gao:1999ar,Gerasyuta:2002kq,Lipkin:1998pb}.

In 2003, LEPS announced the observation of $\Theta^{+}(1540)$ with strangeness $S=+1$ \cite{Nakano:2003qx}, which has the $uudd\bar{s}$ component. Of course, $\Theta^{+}(1540)$ stimulated extensive studies of pentaquark  \cite{Rosner:2003ia,Chen:2016qju,Karliner:2017qhf,Liu:2019zoy}. However, some high precision experiments such as BES \cite{Bai:2004gk}, BaBar \cite{Aubert:2005qi}, Belle \cite{Wang:2005fc} and CDF \cite{Litvintsev:2004yw} did not confirmed the existence of $\Theta^{+}(1540)$. Facing this situation, one realized again that our understanding of non-perturbative behavior of quantum chromodynamics (QCD) is still absent, which is a lesson of $\Theta^{+}(1540)$.

Obviously, it is not the end of the exploration of exotic hadronic matter. With the accumulation of experimental data, more and more charmoniumlike $XYZ$ states have been discovered since 2013, which again inspired theorist's extensive interest in investigating exotic hadronic states \cite{Guo:2014iya,Esposito:2014rxa,Ali:2017jda,Brambilla:2019esw,Hosaka:2016pey,Richard:2016eis,Lebed:2016hpi,Esposito:2016noz}. Especially, in 2015, the LHCb Collaboration measured the $\Lambda_{b}^{0} \rightarrow J/\psi K^{-}p$ decay and observed two hidden-charm pentaquark-like resonances $P_{c}(4380)$ and $P_{c}(4450)$ in the $J/\psi p$ invariant mass spectrum, which indicates that they have a minimal quark content of $uudc\bar{c}$ \cite{Aaij:2015tga,Aaij:2016phn}.
In 2019, LHCb found three narrow $P_c$ structures in the ${J/\psi}p$ invariant mass spectrum of $\Lambda_{b}{\to}{J/\psi}K^{-}p$ \cite{Aaij:2019vzc}, where this new measurement shows that
$P_{c}(4450)$ is actually composed of two substructures, $P_{c}(4440)$ and $P_{c}(4457)$ with $5.4\sigma$ significance. The characteristic mass spectrum of $P_c$
provides the strong evidence of existence of hidden-charm molecular pentaquarks \cite{Chen:2019asm,Wang:2019ato,He:2019rva,Liu:2019zvb,Burns:2019iih}.

At present, searching for exotic hadronic states is still full of challenge and opportunity. As theorists, we should provide valuable prediction, which requires us to
continue to find some crucial hint for exotic hadronic states.

\begin{figure}[htpb]
\includegraphics[width=240pt]{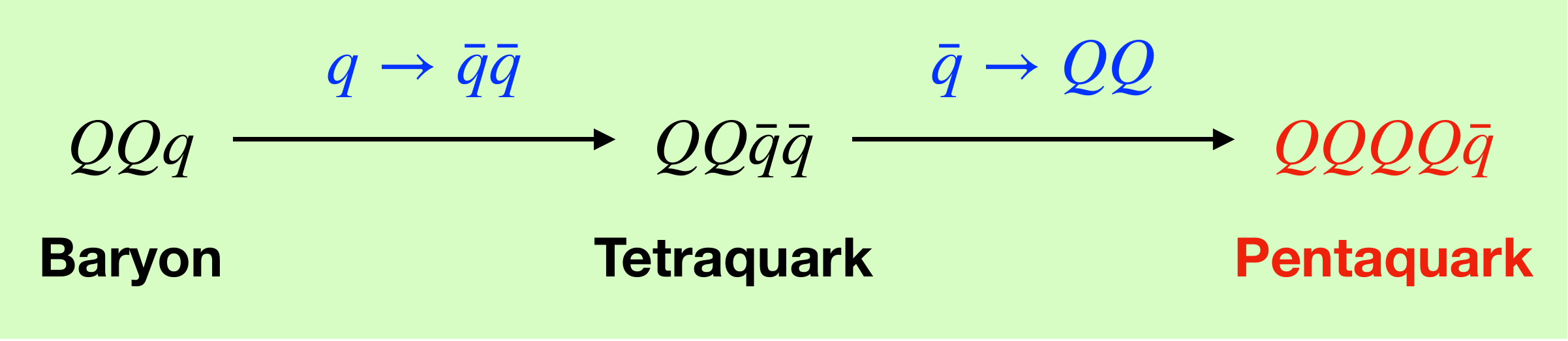}
\caption{Evolution of the $QQQQ\bar q$ pentaquark from double heavy tetraquark and baryon.}\label{aa}
\end{figure}

We may borrow some idea of proposing stable tetraquark state with $QQ\bar{q}\bar{q}$ configuration.
Stimulated by the observation of double charm baryon $\Xi_{cc}^{++}(3620)$ \cite{Aaij:2017ueg}, Refs. \cite{Luo:2017eub,Karliner:2017qjm,Eichten:2017ffp} studied the possible stable tetraquark state with the $QQ\bar{q}\bar{q}$ configuration. Here, the $QQ\bar{q}\bar{q}$ configuration can be obtained by replacing light quark $q$ of double heavy baryon $QQq$ with $\bar q\bar{q}$ pair since the color structure of $q$ and $\bar q\bar{q}$ can be the same. Along this line, we may continue to replace $\bar{q}$ of $QQ\bar{q}\bar{q}$ with a $QQ$ pair and get the $QQQQ\bar{q}$ configuration, which is a typical pentaquark configuration (see Fig. \ref{aa}).

As a new type of pentaquark, the $QQQQ\bar{q}$ pentaquark
draw our attention to explore its mass spectrum and decay behavior. This information is valuable to future experimental search for the $QQQQ\bar{q}$ pentaquark. In this work,
we systematically present the mass spectrum of the S-wave $QQQQ\bar{q}$ system within the framework of Chromomagnetic Interaction (CMI) model \cite{DeRujula:1975qlm}, which has been widely adopted to study the mass spectra of multiquark systems
\cite{Luo:2017eub,Wu:2016gas,Wu:2018xdi,Chen:2016ont,Wu:2016vtq,Liu:2016ogz,Wu:2017weo,Zhou:2018pcv,
Li:2018vhp,An:2019idk,Cheng:2020irt,Cheng:2019obk,Hogaasen:2013nca,Weng:2018mmf,
Weng:2019ynva,Weng:2020jao,Zhao:2014qva,Cheng:2020nho,An:2020jix}.
Moreover, in this work we estimate the ratios of the possible decays of the $QQQQ\bar{q}$ pentaquarks, which are crucial for hunting this kind of pentaquark.

This paper is organized as follows. After introduction, the adopted CMI model will be introduced in In Sec. \ref{sec2}.
In Sec. \ref{sec4}, we present the mass spectra of S-wave $QQQQ\bar{q}$ pentaquarks and estimate the ratios of possible strong decay widths. Finally,
a short summary is followed in Sec. \ref{sec5}.

\section{The chromomagnetic Interaction Model}\label{sec2}

The effective Hamiltonian involved in the estimated mass spectrum of these discussed pentaquarks
\begin{eqnarray}\label{Eq1}
H&=&\sum_im_i+H_{\textrm{CMI}} \nonumber \\
&=&\sum_im_i-\sum_{i<j}C_{ij} \vec\lambda_i\cdot \vec\lambda_j \vec\sigma_i\cdot\vec\sigma_j,
\end{eqnarray}
where $m_i$ denotes the effective mass of the $i$-th constituent quark when considering these effects from kinetic energy, color confinement, and so on.
$H_{\rm CMI}$, as the chromomagnetic interaction Hamiltonian, is composed of the Pauli matrices $\sigma_{i}$  and the Gell-Mann matrices $\lambda_i$. Here,
$\lambda_i$ should be replaced by $-\lambda_i^{*}$ for antiquark, and $C_{ij}$ denotes the effective coupling constant between the $i$-th quark and $j$-th quark, which depends on the quark masses and the ground particle's spatial wave function. As input, $C_{ij}$ is fixed by the involved hadron masses.

Usually, there exists overestimate to the predicted hadron masses by Eq. (\ref{Eq1})
\cite{Wu:2016vtq, Wu:2016gas, Chen:2016ont, Wu:2017weo, Luo:2017eub, Zhou:2018pcv, Li:2018vhp, Wu:2018xdi,An:2019idk,Liu:2016ogz,Cheng:2020irt,Cheng:2019obk}. Additionally, the effective mass $m_{i}$ in Eq. \ref{Eq1} does not incorporate attraction sufficiently.
The mass of the pentaquark state reads as
\begin{eqnarray}
\label{Eq11}
M=M_{ref}-\langle H_{\rm CMI}\rangle_{ref}+\langle H_{\rm CMI}\rangle,
\end{eqnarray}
where $M_{ref}$, $\langle H_{\rm CMI}\rangle_{ref}$, and $\langle H_{\rm CMI}\rangle$ are the physical mass of the reference system, the corresponding CMI eigenvalue, and the CMI eigenvalue for the discussed multi-quark state, respectively.
For $M_{ref}$, we can use the threshold of a reference baryon-meson system whose quark content is the same as the considered pentaquark states.
For example, for the $nncc\bar{c}$ system, we can use $(M_{\Xi_{cc}}+M_{D})-(\langle H_{\rm CMI}\rangle_{\Xi_{cc}}+\langle H_{\rm CMI}\rangle_{D})+\langle H_{\rm CMI}\rangle_{nncc\bar{c}}$
or $(M_{\Sigma_{c}}+M_{J/\psi})-(\langle H_{\rm CMI}\rangle_{\Sigma_{c}}+\langle H_{\rm CMI}\rangle_{J/\psi})+\langle H_{\rm CMI}\rangle_{nncc\bar{c}}$ to obtain the mass spectra.
All possible used $M_{ref}$ values are shown in Table \ref{comp}
\footnote{Since some of the heavy flavor baryons are not-yet observed, we adopt the theoretical results in Ref. \cite{Weng:2018mmf} as input.
}
With this treatment, the dynamical effects that are not incorporated in the original approach can be compensated \cite{Zhou:2018pcv}. This approach of determining the masses of pentaquark states is named as the reference mass scheme here.

Alternatively, a color-electric term can be introduced here \cite{Hogaasen:2013nca,Liu:2019zoy,Weng:2018mmf,Weng:2019ynva,Weng:2020jao,An:2020jix}
\begin{eqnarray}\label{Eq2}
H_{\rm CEI}=-\sum_{i<j}A_{ij} \vec\lambda_i\cdot \vec\lambda_j.
\end{eqnarray}

With deduction
\begin{eqnarray}
&&\sum_{i<j}(m^{0}_{i}+m^{0}_{j})\vec\lambda_i\cdot \vec\lambda_j\nonumber\\
&&=\frac{1}{2}\sum_{i,j}(m^{0}_{i}+m^{0}_{j})\vec\lambda_i\cdot \vec\lambda_j-\sum_{i}m^{0}_{i}(\vec\lambda_i)^{2}\nonumber\\
&&=(\sum_{i}m^{0}_{i}\vec\lambda_i)\cdot\left(\sum_{i}\vec\lambda_i\right)-\frac{16}{3}\sum_{i}m^{0}_{i},
\end{eqnarray}
where the color operator $\sum_{i}\vec\lambda_i$ nullifies any colorless physical
state, we have the Hamiltonian of the modified CMI model
\begin{eqnarray}\label{Eq3}
H&=&\sum_im^{0}_i+H_{\rm CEI}+H_{\rm CMI}\nonumber\\
&=&\sum_im^{0}_i-\sum_{i<j}A_{ij} \vec\lambda_i\cdot \vec\lambda_j-\sum_{i<j}v_{ij} \vec\lambda_i\cdot \vec\lambda_j \vec\sigma_i\cdot\vec\sigma_j, \nonumber\\
&=&-\frac{3}{4}(-\frac{4}{3}\sum_im^{0}_i)-\sum_{i<j}A_{ij} \vec\lambda_i\cdot \vec\lambda_j\nonumber\\&&-\sum_{i<j}v_{ij} \vec\lambda_i\cdot \vec\lambda_j \vec\sigma_i\cdot\vec\sigma_j, \nonumber\\
&=&-\frac{3}{4}\sum_{i<j}[\frac{1}{4}(m^{0}_{i}+m^{0}_{j})+\frac{4}{3}A_{ij}] \vec\lambda_i\cdot \vec\lambda_j\nonumber\\&&-\sum_{i<j}v_{ij} \vec\lambda_i\cdot \vec\lambda_j \vec\sigma_i\cdot\vec\sigma_j, \nonumber\\
&=&-\frac{3}{4}\sum_{i<j}m_{ij}V^{\rm C}_{ij}-\sum_{i<j}v_{ij}V^{\rm CMI}_{ij}.
\end{eqnarray}
In the above expression, $V^{\rm C}_{ij}=\vec\lambda_i\cdot \vec\lambda_j$ and $ V^{\rm CMI}_{ij}=\vec\lambda_i\cdot \vec\lambda_j \vec\sigma_i\cdot\vec\sigma_j$
denote the color and color-magnetic interactions between quarks, respectively.
A new mass parameter of quark pair should be defined, i.e.,
\begin{eqnarray}
m_{ij}=\frac{1}{4}(m^{0}_{i}+m^{0}_{j})+\frac{4}{3}A_{ij}.
\end{eqnarray}
By these conventional hadron masses (see Table \ref{comp}), we may fix the parameters $m_{ij}$ and $v_{ij}$ (see Table \ref{parameter2}). Interested readers can refer to Refs. \cite{Weng:2019ynva,Weng:2018mmf,Liu:2019zoy,Weng:2020jao} for more details. With this modified CMI model scheme, we can give mass spectrum of the $QQQQ\bar{q}$ pentaquarks.

\renewcommand\arraystretch{1.5}
\begin{table}[htbp]
\caption{The adopted masses of conventional hadrons for determining parameters (in units of MeV) \cite{Tanabashi:2018oca,Weng:2018mmf}.
}\label{comp}
\begin{tabular}{crc|crc}
\bottomrule[1.5pt]
\bottomrule[0.5pt]
hadrons& $I(J^P)$ &Mass & hadrons& $I(J^P)$ &Mass \\
\bottomrule[0.7pt]
$D$&$1/2(0^{-})$&1869.6&$D^{*}$&$1/2(1^{-})$&2010.3\\
$D_{s}$&$0(0^{-})$&1968.3&$D^{*}_{s}$&$0(1^{-})$&2112.2\\
$B$&$1/2(0^{-})$&5279.3&$B^{*}$&$1/2(1^{-})$&5324.7\\
$B_{s}$&$0(0^{-})$&5366.8&$B^{*}_{s}$&$0(1^{-})$&5415.4\\
$\Xi_{cc}$&$1/2(1/2^{+})$&3621.4&$\Xi^{*}_{cc}$&$1/2(3/2^{+})$&(3696.1)\\
$\Omega_{cc}$&$0(1/2^{+})$&(3731.8)&$\Omega^{*}_{cc}$&$0(3/2^{+})$&(3802.4)\\
$\Xi_{cb}$&$1/2(1/2^{+})$&(6922.3)&$\Omega_{cb}$&$0(1/2^{+})$&(7010.7)\\
$\Xi'_{cb}$&$1/2(1/2^{+})$&(6947.9)&$\Omega'_{cb}$&$0(1/2^{+})$&(7047.0)\\
$\Xi^{*}_{cb}$&$1/2(3/2^{+})$&(6973.2)&$\Omega^{*}_{cb}$&$0(3/2^{+})$&(7065.7)\\
$\Xi_{bb}$&$1/2(1/2^{+})$&(10168.9)&$\Xi^{*}_{bb}$&$1/2(3/2^{+})$&(10188.8)\\
$\Omega_{bb}$&$0(1/2^{+})$&(10259.0)&$\Omega^{*}_{bb}$&$0(3/2^{+})$&(10267.5)\\
$\Omega_{ccc}$&$0(3/2^{+})$&(4785.6)&$\Omega_{bbb}$&$0(3/2^{+})$&(14309.7)\\
$\Omega_{ccb}$&$0(1/2^{+})$&(7990.3)&$\Omega_{ccb}^{*}$&$0(3/2^{+})$&(8021.8)\\
$\Omega_{bbc}$&$0(1/2^{+})$&(11165.0)&$\Omega_{bbc}^{*}$&$0(3/2^{+})$&(11196.4)\\
\bottomrule[0.5pt]
\midrule[1.5pt]
\end{tabular}
\end{table}

\begin{table}[t]
\centering \caption{Coupling parameters for the schemes in units of MeV. Here, $q=n,\,s$ ($n=u$, $d$) and $Q=c,\,b$.
}\label{parameter2}
\renewcommand\arraystretch{1.5}
\begin{tabular}{ccccccc}
\bottomrule[1.5pt]
\bottomrule[0.5pt]
\multicolumn{7}{c}{The reference mass scheme}\\
$C_{cc}$&$C_{bb}$&$C_{cb}$&$C_{c\bar{n}}$&$C_{c\bar{s}}$&$C_{b\bar{n}}$&$C_{b\bar{s}}$\\
3.3&1.8&2.0&6.6&6.7&2.1&2.3\\
\bottomrule[1.0pt]
\multicolumn{7}{c}{The modified CMI model scheme}\\
$m_{cc}$&$m_{bb}$&$m_{cb}$&$m_{c\bar{n}}$&$m_{c\bar{s}}$&$m_{b\bar{n}}$&$m_{b\bar{s}}$\\
792.9&2382.4&1604.0&493.3&519.0&1328.3&1350.8\\
\bottomrule[0.5pt]
$v_{cc}$&$v_{bb}$&$v_{cb}$&$v_{c\bar{n}}$&$v_{c\bar{s}}$&$v_{b\bar{n}}$&$v_{b\bar{s}}$\\
3.5&1.9&2.0&6.6&6.7&2.1&2.3\\
\bottomrule[0.5pt]
\bottomrule[1.5pt]
\end{tabular}
\end{table}

When calculating the mass of the pentaquark states, we need the information of the total wave function, which is composed of space, flavor, color, and spin wave functions, i.e.,
\begin{eqnarray}
\psi_{\textrm{tot}}=\psi_{space}\otimes\psi_{flavor}\otimes\psi_{color}\otimes\psi_{spin}.
\end{eqnarray}
Since we only focus on the low-lying $S$-wave pentaquark states, the symmetrical constraint from spatial pentaquark wave function is trivial. For the discussed $QQQQ\bar{q}$ pentaquarks, their $\psi_{flavor}\otimes\psi_{color}\otimes\psi_{spin}$ wave functions of pentaquark states should be fully antisymmetric when exchanging identical quarks.
All the possible flavor combinations for the $QQQQ\bar{q}$ pentaquark system are shown in Table \ref{flavor1}, by which we further determine $\psi_{flavor}\otimes\psi_{color}\otimes\psi_{spin}$ wave functions, which satisfy $\{1234\}$, $\{123\}$, and $\{12\}\{34\}$ symmetry. Here, we use the notation $\{1234\}$ to label that the 1st, 2nd, 3rd, and 4th quarks have antisymmetry property, which will be applied in the following discussions. For identifying the pentaquark configuration with certain exchanging symmetry, we use the approach of Young diagram and Young tableau, which represents the irreducible bases of the permutation group.

\begin{table}[t]
\centering \caption{All possible flavor combinations for the $QQQQ\bar{q}$ pentaquark system with $q=n,\,s$ ($n=u$, $d$) and $Q=c,\,b$.
}\label{flavor1}
\renewcommand\arraystretch{1.4}
\begin{tabular}{p{1.3cm}|p{1.3cm}p{1.3cm}p{1.3cm}p{1.3cm}p{1.3cm}}
\bottomrule[1.5pt]
\bottomrule[0.5pt]
System&\multicolumn{5}{c}{Flavor combinations}\\
\bottomrule[0.5pt]
\multirow{2}*{$QQQQ\bar{q}$}&$cccc\bar{n}$&$cccb\bar{n}$&$ccbb\bar{n}$&$bbbc\bar{n}$&$bbbb\bar{n}$\\
&$cccc\bar{s}$&$cccb\bar{s}$&$ccbb\bar{s}$&$bbbc\bar{s}$&$bbbb\bar{s}$\\
\bottomrule[0.5pt]
\bottomrule[1.5pt]
\end{tabular}
\end{table}

The color wave functions for $QQQQ\bar{q}$ pentaquark system are expressed a direct product
\begin{eqnarray}
\left[3_c\otimes3_c\otimes3_c\otimes3_c\right]\otimes\bar{3}_c.\label{colorproduct}
\end{eqnarray}
Due to the requirement of color confinement,
the color wave function must be a singlet.
Therefore, the four heavy quarks should be in the color triplet states, and the corresponding partition [211] reads as
\begin{align}
&\begin{tabular}{|c|c|}
\hline
           1 &  2    \\
\cline{1-2}
\multicolumn{1}{|c|}{3} \\
\cline{1-1}
\multicolumn{1}{|c|}{4}  \\
\cline{1-1}
\end{tabular}
=\left\{\left(12\right)_6\left(34\right)_{\bar{3}}\right\}_3,
\quad
\begin{tabular}{|c|c|}
\hline
           1 &  3    \\
\cline{1-2}
\multicolumn{1}{|c|}{2} \\
\cline{1-1}
\multicolumn{1}{|c|}{4}  \\
\cline{1-1}
\end{tabular}
=\left\{\left(12\right)_{\bar{3}} 34\right\}_3,
\quad
\nonumber \\
&\begin{tabular}{|c|c|}
\hline
   1 &  4   \\
\cline{1-2}
\multicolumn{1}{|c|}{2} \\
\cline{1-1}
\multicolumn{1}{|c|}{3} \\
\cline{1-1}
\end{tabular}
=\left\{\left(123\right)_14\right\}_3.
\label{eq-color1}
\end{align}
Here, the subscript labels the irreducible representation of $\rm SU(3)$.
Then, by combining the antitriplet from light anti-quark with the deduced three color triplets in Eq. (\ref{eq-color1}), we obtain
three color singlets for all the studied pentquark systems
\begin{eqnarray}
\label{eq-color2}
\vert C_1 \rangle&=&\vert \left [\left(12\right)_{6}\left(34\right)_{\bar{3}}\right ]_{3} (\bar{5})_{\bar{3}}\rangle =
\begin{tabular}{|c|c|}
\hline
1 &  2    \\
\cline{1-2}
\multicolumn{1}{|c|}{3} \\
\cline{1-1}
\multicolumn{1}{|c|}{4}  \\
\cline{1-1}
\end{tabular}_3
\otimes(\bar{5})_{\bar{3}}\nonumber \\
&=&\frac{1}{4\sqrt{3}}\Big[(2bbgr-2bbrg+gbrb-gbbr+bgrb-bgbr \nonumber\\
&&-rbgb+rbbg-brgb+brbg)\bar{b}+(2rrbg-2rrgb \nonumber\\
&&+rgrb-rgbr+grrb-grbr+rbgr-rbrg+brgr\nonumber\\
&&-brrg)\bar{r}+(2ggrb-2ggbr-rggb+rgbg-grgb \nonumber\\
&&+grbg+gbgr-gbrg+bggr-bgrg)\bar{g}\Big],
\end{eqnarray}

\begin{eqnarray}
\label{eq-color3}
\vert C_2 \rangle&=&\vert \left [\left(12\right)_{\bar{3}}34\right ]_{3} (\bar{5})_{\bar{3}}\rangle =
\begin{tabular}{|c|c|}
\hline
  1 &  3    \\
\cline{1-2}
\multicolumn{1}{|c|}{2} \\
\cline{1-1}
\multicolumn{1}{|c|}{4}  \\
\cline{1-1}
\end{tabular}_3
\otimes(\bar{5})_{\bar{3}} \nonumber\\
&=&\frac{1}{12}\Big[(3bgbr-3gbbr-3brbg+3rbbg-rbgb+gbrb \nonumber\\
&&+2grbb+brgb-bgrb-2rgbb)\bar{b}+(3grrb-3rgrb \nonumber\\
&&-3brrg+3rbrg-rbgr-2gbrr+2bgrr-grbr  \nonumber \\
&&+rgbr+brgr)\bar{r}+(3grgb-3rggb+3bggr-3gbgr\nonumber \\
&&-grbg+rgbg+2rbgg-2brgg+gbrg-bgrg)\bar{g}\Big], \nonumber\\
\end{eqnarray}
and

\begin{eqnarray}
\label{eq-color4}
\vert C_3 \rangle&=&\vert \left [\left(123\right)_{1}4\right ]_{3} (\bar{5})_{\bar{3}}\rangle =
\begin{tabular}{|c|c|}
\hline
1 &  4    \\
\cline{1-2}
\multicolumn{1}{|c|}{2} \\
\cline{1-1}
\multicolumn{1}{|c|}{3}  \\
\cline{1-1}
\end{tabular}_3
\otimes(\bar{5})_{\bar{3}}\nonumber\\
&=&\frac{1}{3\sqrt{2}}\Big[(grbb-rgbb+rbgb-brgb+bgrb-gbrb)\bar{b}\nonumber\\
&&+(grbr-rgbr+rbgr-brgr+bgrr-gbrr)\bar{r}\nonumber \\
&&+(grbg-rgbg+rbgg-brgg+bgrg-gbrg)\bar{g}\Big]. \nonumber\\
\end{eqnarray}

Next, we discuss spin wave function in spin space.
For the $QQQQ\bar q$ pentaquark system with total spin $J=5/2$, the spin state can be represented by partition [5], i.e.,
\begin{align}
\label{eq-spin1}
\begin{tabular}{|c|c|c|c|c|}
\hline
1&2&3&4&5   \\
\cline{1-5}
\end{tabular}_{\begin{tabular}{|c|} \multicolumn{1}{c}{$S_1$}\end{tabular}}.\\ \nonumber
\end{align}
The spin wave functions for the $QQQQ\bar q$ pentaquark with $J=3/2$ are represented in partition [41] as
\begin{align}
\label{eq-spin2}
&\begin{tabular}{|c|c|c|c|}
\hline
1&2&3&4  \\
\cline{1-4}
\multicolumn{1}{|c|}{5} \\
\cline{1-1}
\end{tabular}_{\begin{tabular}{|c|} \multicolumn{1}{c}{$S_1$}\end{tabular}},
\begin{tabular}{|c|c|c|c|}
\hline
1&2&3&5  \\
\cline{1-4}
\multicolumn{1}{|c|}{4} \\
\cline{1-1}
\end{tabular}_{\begin{tabular}{|c|} \multicolumn{1}{c}{$S_2$}\end{tabular}},
\begin{tabular}{|c|c|c|c|}
\hline
1&2&4&5  \\
\cline{1-4}
\multicolumn{1}{|c|}{3} \\
\cline{1-1}
\end{tabular}_{\begin{tabular}{|c|} \multicolumn{1}{c}{$S_3$}\end{tabular}},
\begin{tabular}{|c|c|c|c|}
\hline
1&3&4&5  \\
\cline{1-4}
\multicolumn{1}{|c|}{2} \\
\cline{1-1}
\end{tabular}_{\begin{tabular}{|c|} \multicolumn{1}{c}{$S_4$}\end{tabular}}. \nonumber \\
\end{align}
With the similar method, one obtains the spin wave functions for the $QQQQ\bar q$ pentaquark states with $J=1/2$, which can be represented in partition [32] as
\begin{align}
\label{eq-spin3}
&\begin{tabular}{|c|c|c|}
\hline
1&2&3  \\
\cline{1-3}
4&5 \\
\cline{1-2}
\end{tabular}_{\begin{tabular}{|c|} \multicolumn{1}{c}{$S_1$}\end{tabular}},
\begin{tabular}{|c|c|c|}
\hline
1&2&4  \\
\cline{1-3}
3&5 \\
\cline{1-2}
\end{tabular}_{\begin{tabular}{|c|} \multicolumn{1}{c}{$S_2$}\end{tabular}},
\begin{tabular}{|c|c|c|}
\hline
1&3&4  \\
\cline{1-3}
2&5 \\
\cline{1-2}
\end{tabular}_{\begin{tabular}{|c|} \multicolumn{1}{c}{$S_3$}\end{tabular}},
\begin{tabular}{|c|c|c|}
\hline
1&2&5  \\
\cline{1-3}
3&4 \\
\cline{1-2}
\end{tabular}_{\begin{tabular}{|c|} \multicolumn{1}{c}{$S_4$}\end{tabular}},
\begin{tabular}{|c|c|c|}
\hline
1&3&5  \\
\cline{1-3}
2&4 \\
\cline{1-2}
\end{tabular}_{\begin{tabular}{|c|} \multicolumn{1}{c}{$S_5$}\end{tabular}}.\\ \nonumber
\end{align}
Since the particle 5 corresponds to a light antiquark, we can isolate this antiquark and discuss the symmetry property of the first four heavy quarks 1, 2, 3, and 4 in $\psi_{color}\otimes\psi_{spin}$ space.

When the antiquark 5 is separated from the spin wave functions,
the spin states represented in Young tableaux without the antiquark 5 can be directly obtained from Eqs. (\ref{eq-spin1}-\ref{eq-spin3}) as
\begin{align}
\label{eq-spin4}
J=\frac52: \quad
&\begin{tabular}{|c|c|c|c|}
\hline
1&2&3&4   \\
\cline{1-4}
\end{tabular}_{\begin{tabular}{|c|} \multicolumn{1}{c}{$S_1$}\end{tabular}} ,
\nonumber
\\
J=\frac32:  \quad
&\begin{tabular}{|c|c|c|c|}
\hline
1&2&3&4  \\
\cline{1-4}
\end{tabular}_{\begin{tabular}{|c|} \multicolumn{1}{c}{$S_1$}\end{tabular}} ,
\begin{tabular}{|c|c|c|}
\hline
1&2&3 \\
\cline{1-3}
\multicolumn{1}{|c|}{4} \\
\cline{1-1}
\end{tabular}_{\begin{tabular}{|c|} \multicolumn{1}{c}{$S_2$}\end{tabular}},
\begin{tabular}{|c|c|c|c|}
\hline
1&2&4  \\
\cline{1-3}
\multicolumn{1}{|c|}{3} \\
\cline{1-1}
\end{tabular}_{\begin{tabular}{|c|} \multicolumn{1}{c}{$S_3$}\end{tabular}},
\begin{tabular}{|c|c|c|c|}
\hline
1&3&4  \\
\cline{1-3}
\multicolumn{1}{|c|}{2} \\
\cline{1-1}
\end{tabular}_{\begin{tabular}{|c|} \multicolumn{1}{c}{$S_4$}\end{tabular}},
\nonumber
\\
J=\frac12:  \quad
&\begin{tabular}{|c|c|c|}
\hline
1&2&3  \\
\cline{1-3}
4 \\
\cline{1-1}
\end{tabular}_{\begin{tabular}{|c|} \multicolumn{1}{c}{$S_1$}\end{tabular}},
\begin{tabular}{|c|c|c|}
\hline
1&2&4  \\
\cline{1-3}
3 \\
\cline{1-1}
\end{tabular}_{\begin{tabular}{|c|} \multicolumn{1}{c}{$S_2$}\end{tabular}},
\begin{tabular}{|c|c|c|}
\hline
1&3&4  \\
\cline{1-3}
2 \\
\cline{1-1}
\end{tabular}_{\begin{tabular}{|c|} \multicolumn{1}{c}{$S_3$}\end{tabular}},
\begin{tabular}{|c|c|}
\hline
1&2  \\
\cline{1-2}
3&4 \\
\cline{1-2}
\end{tabular}_{\begin{tabular}{|c|} \multicolumn{1}{c}{$S_4$}\end{tabular}},
\begin{tabular}{|c|c|}
\hline
1&3 \\
\cline{1-2}
2&4 \\
\cline{1-2}
\end{tabular}_{\begin{tabular}{|c|} \multicolumn{1}{c}{$S_5$}\end{tabular}}.\nonumber \\
\end{align}
In Eq. (\ref{eq-spin4}), the spin states can be identified with the Young-Yamanouchi basis vectors for partitions [4], [31], and [22].

For constructing the $\psi_{flavor}\otimes\psi_{color}\otimes\psi_{spin}$ wave functions of $QQQQ\bar{q}$ pentaquark system, we should combine the  partition [211] of the color singlets in Eq. (\ref{eq-color1}) with  partitions [4], [31], [22] of the spin states in Eq. (\ref{eq-spin4}) by the inner product of the permutation group. Thus, the $\psi_{color}\otimes\psi_{spin}$ wave functions with a certain symmetry can be constructed.
We get the corresponding Young diagram representations of $\psi_{color}\otimes\psi_{spin}$ bases \cite{Itzykson:1965hk,Stancu:1999qr,Park:2017jbn}
\begin{flalign}\label{colorspin}
\includegraphics[width=\columnwidth]{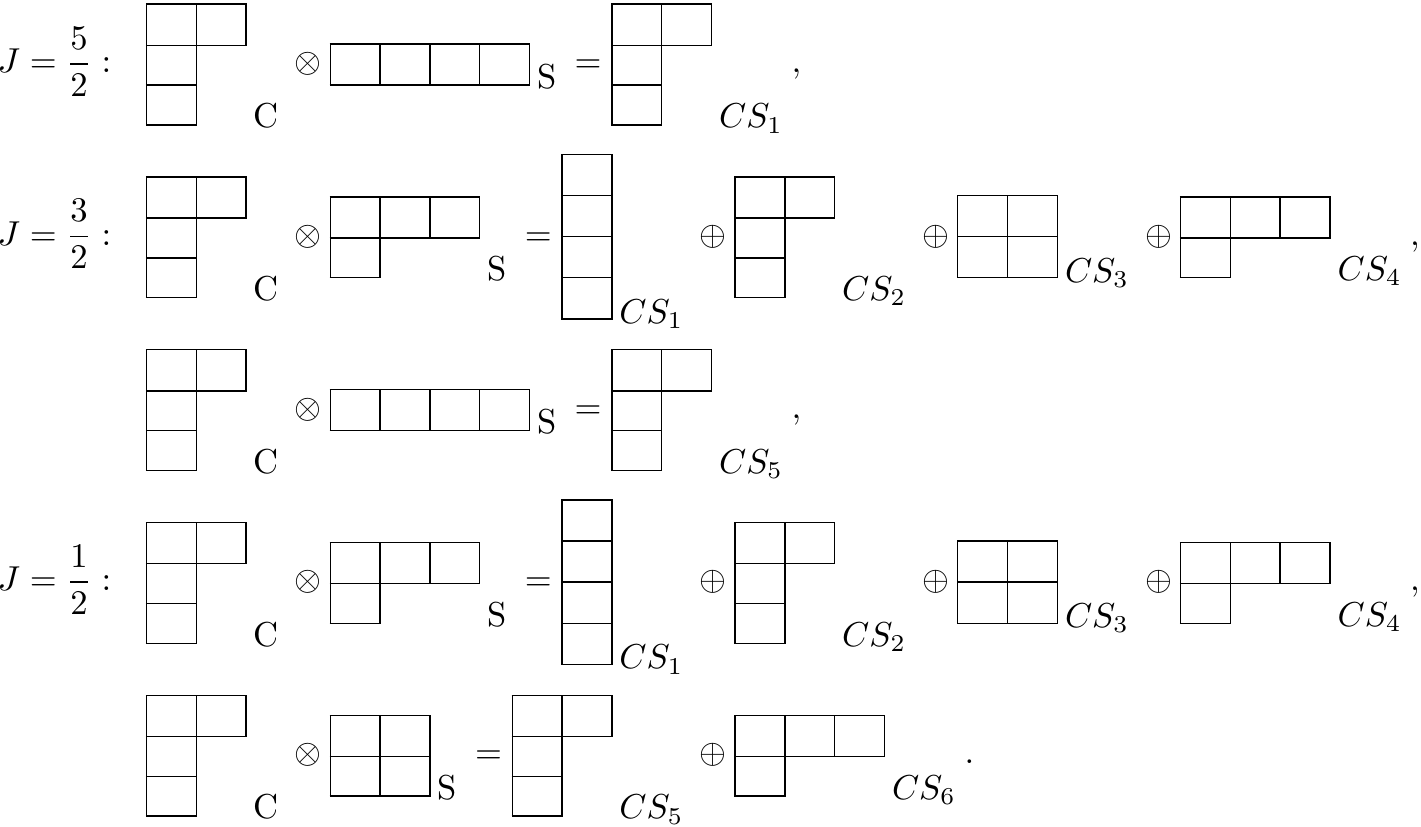}
\end{flalign}

By using the Clebsch-Gordan (CG) coefficient of the permutation group $S_{n}$, one obtains the coupling scheme designed to construct the $\psi_{color}\otimes\psi_{spin}$ states.
Here, any CG coefficient of $S_{n}$ can be factorized into an isoscalar factor, called $K$ matrix, and the Clebsch-Gordan (CG) coefficient of $S_{n-1}$ \cite{Stancu:1999qr}.
The expression of Clebsch-Gordan (CG) coefficient of $S_{n}$ reads as
\begin{align}
&S([f^{\prime}]p^{\prime}q^{\prime}y^{\prime}[f^{\prime\prime}]p^{\prime\prime}q^{\prime\prime}y^{\prime\prime}\vert[f]pqy)=\nonumber\\
&K([f^{\prime}]p^{\prime}[f^{\prime\prime}]p^{\prime\prime}\vert[f]p)
S([f^{\prime}_{p^{\prime}}]q^{\prime}y^{\prime}[f^{\prime\prime}_{p^{\prime\prime}}]q^{\prime\prime}y^{\prime\prime}\vert[f_p]qy).
\label{K-matrix}
\end{align}
Here, $S$ in the left-hand (right-hand) side denotes a CG coefficient of $S_n$ ($S_{n-1}$) and
the $[f]$ ($[f_p]$) is a Young tableau of $S_{n}$ ($S_{n-1}$) with $[f]pqy$ as a specific Young-Yamanouchi basis vector. And, $p$ ($q$) is the row of the $n$ $(n-1)$-th particle in the Young-Yamanouchi basis vector, while $y$ is the distribution of the $n-2$ remaining particles. In fact, a similar application is also used in Refs. \cite{Park:2017jbn,Park:2016mez,Park:2016cmg,Park:2015nha,Park:2018oib}.
According to the isoscalar factors for $S_{3}$ and $S_{4}$ in Tables 6.2 and 6.3 of Ref. \cite{Stancu:1999qr}, the corresponding Clebsch-Gordan (CG) coefficient of $S_{4}$ can be obtained. The corresponding Young-Yamanouchi basis vector obtained from the $\psi_{color}\otimes\psi_{spin}$ coupling (see Eq.(\ref{colorspin})) are collected in Eq. (\ref{colorspin1}) of Appendix. \ref{sec10}.

We can combine the flavor wave functions with the $\psi_{color}\otimes\psi_{spin}$ wave functions (see Eq. \ref{colorspin1}) of $QQQQ\bar q$ pentaquark states to deduce the symmetry allowed $\psi_{flavor}\otimes\psi_{color}\otimes\psi_{spin}$ pentaquark wave functions.
In Table. \ref{flavor} of Appendix. \ref{sec10}, all the symmetrically allowed Young-Yamanouchi basis vector for the different flavor wave functions are listed.

By constructing all the possible $\psi_{flavor}\otimes\psi_{color}\otimes\psi_{spin}$ bases satisfied for $\{1234\}$, $\{123\}$, and $\{12\}\{34\}$ symmetry,
we can calculate the CMI matrices for the studied pentaquark states.
Here, we only present the expressions of CMI Hamiltonians for the $cccc\bar{n}$, $cccb\bar{n}$, and $ccbb\bar{n}$ pentaquark subsystems in Table \ref{nnnsQ}  of Appendix. \ref{sec10}.
According to their similar symmetry properties, the expressions of CMI matrices for other pentaquark subsystems can be deduced for those of the the $cccc\bar{c}$, $cccb\bar{c}$, and $ccbb\bar{c}$ pentaquark subsystems.

\section{Mass Spectra and Decay Behaviors}\label{sec4}
As shown in Table \ref{flavor1}, according to symmetry properties, we can divide the $QQQQ\bar{q}$ pentaquark system into three groups: \\
\begin{enumerate}
\item[A.] The $cccc\bar{q}$ and $bbbb\bar{q}$ pentaquark subsystems;
\item[B.]  The $cccb\bar{q}$ and $bbbc\bar{q}$ pentaquark subsystems;
\item[C.]  The $ccbb\bar{q}$ pentaquark subsystem.
\end{enumerate}
For the $cccc\bar{q}$ and $bbbb\bar{q}$ subsystems, the wave functions should be to antisymmetric for the exchange between the 12, 13, 14, 23, 24, or 34 particles.
The $cccb\bar{q}$ and $bbbc\bar{q}$ wave functions are antisymmetric when exchanging the 12, 13, or 23 particles.
However, for the $ccbb\bar{q}$ subsystem, the antisymmetry is considered only for exchanging the 12 or 34 particles.
The fewer restrictions lead to more allowed wave functions.
Therefore, the number of basis states in Table \ref{flavor} increases from $cccc\bar{n}$ to $cccb\bar{n}$ and next to $ccbb\bar{n}$ due to the Pauli principle.
In the following, we will discuss the mass spectra and strong decay properties of $QQQQ\bar{q}$ pentaquark system group by group.
All of them are explicitly exotic states.
If such pentaquark states could be observed in experiment, its pentaquark state nature could be easily identified.
For simplicity, we use $\rm P_{content}$(Mass, $I$, $J^{P}$) to label a particular pentaquark state.

\begin{table*}[t]
\centering \caption{The estimated masses for the $QQQQ\bar{q}$ $(Q=c,b; q=n,s; n=u,d)$ system in units of MeV. The eigenvalues of $H_{\textrm{CMI}}$ matrix are listed in the second column.
The corresponding masses in the reference mass scheme are listed in third and/or fourth columns. The masses with the modified CMI model scheme are presented in the last column.}
\label{mass-QQQQq}
\renewcommand\arraystretch{1.25}
\begin{tabular}{cccc|c|cccc|c}
\bottomrule[1.5pt]
\bottomrule[0.5pt]
&\multicolumn{3}{c|}{reference mass}&{modified CMI}&&\multicolumn{3}{c|}{reference mass}&{modified CMI}\\
&\multicolumn{3}{c|}{scheme}&{model scheme}&&\multicolumn{3}{c|}{scheme}&{model scheme}
\\
\bottomrule[1pt]
\multicolumn{5}{l|}{$cccc\bar{n}$}&\multicolumn{5}{l}{$cccc\bar{s}$}\\
$J^P$&Eigenvalue&($D\Omega_{ccc}$)&&Mass&$J^P$&Eigenvalue&$(D_{s}\Omega_{ccc})$&&Mass\\
\bottomrule[0.7pt]
$\frac{3}{2}^{-}$ &
$26.4$&
$6761$&
&
$6761$&
$\frac{3}{2}^{-}$ &
$25.9$&
$6861$&
&
$6864$\\
$\frac{1}{2}^{-}$ &
$132.0$&
$6867$&
&
$6867$&
$\frac{1}{2}^{-}$ &
133.1&
6968&
&
6972\\
\bottomrule[0.5pt]
\multicolumn{5}{l|}{$bbbb\bar{n}$}&\multicolumn{5}{l}{$bbbb\bar{s}$}\\
$J^P$&Eigenvalue&($B\Omega_{bbb}$)&&Mass&$J^P$&Eigenvalue&$(B_{s}\Omega_{bbb})$&&Mass\\
\bottomrule[0.7pt]
$\frac{3}{2}^{-}$ &
$22.4$&
$19631$&
&
$19647$&
$\frac{3}{2}^{-}$ &
$21.3$&
$19720$&
&
$19736$\\
$\frac{1}{2}^{-}$ &
$56.0$&
$19664$&
&
$19681$&
$\frac{1}{2}^{-}$ &
$58.1$&
$19757$&
&
$19773$\\
\bottomrule[1.0pt]
\multicolumn{5}{l|}{$cccb\bar{n}$}&\multicolumn{5}{l}{$cccb\bar{s}$}\\
$J^P$&Eigenvalue&($B\Omega_{ccc}$)&($D\Omega_{ccb}$)&Mass&$J^P$&Eigenvalue&$(B_{s}\Omega_{ccc})$&$(D_{s}\Omega_{ccb})$&Mass\\
\bottomrule[0.7pt]
$\frac{5}{2}^{-}$ &
$37.6$&
$10110$&
$9816$&
$10110$&
$\frac{5}{2}^{-}$ &
$38.7$&
$10202$&
$9917$&
$10201$\\
$\frac{3}{2}^{-}$ &
$\begin{pmatrix}65.3\\22.0\\-40.6\end{pmatrix}$&
$\begin{pmatrix}10138\\10094\\10032\end{pmatrix}$&
$\begin{pmatrix}9843\\9800\\9738\end{pmatrix}$&
$\begin{pmatrix}10118\\10078\\9961\end{pmatrix}$&
$\frac{3}{2}^{-}$ &
$\begin{pmatrix}65.9\\21.3\\-43.1\end{pmatrix}$&
$\begin{pmatrix}10229\\10184\\10120\end{pmatrix}$&
$\begin{pmatrix}9944\\9900\\9835\end{pmatrix}$&
$\begin{pmatrix}10210\\10168\\10062\end{pmatrix}$\\
$\frac{1}{2}^{-}$ &
$\begin{pmatrix}110.4\\68.0\\-41.3\end{pmatrix}$&
$\begin{pmatrix}10183\\10140\\10031\end{pmatrix}$&
$\begin{pmatrix}9889\\9846\\9737\end{pmatrix}$&
$\begin{pmatrix}10134\\10062\\9946\end{pmatrix}$&
$\frac{1}{2}^{-}$ &
$\begin{pmatrix}111.6\\69.1\\-41.7\end{pmatrix}$&
$\begin{pmatrix}10275\\10232\\10121\end{pmatrix}$&
$\begin{pmatrix}9990\\9948\\9837\end{pmatrix}$&
$\begin{pmatrix}10228\\10166\\10047\end{pmatrix}$\\
\bottomrule[0.5pt]
\multicolumn{5}{l|}{$bbbc\bar{n}$}&\multicolumn{5}{l}{$bbbc\bar{s}$}\\
$J^P$&Eigenvalue&($D\Omega_{bbb}$)&($B\Omega_{bbc}$)&Mass&$J^P$&Eigenvalue&$(D_{s}\Omega_{bbb})$&$(B_{s}\Omega_{bbc})$&Mass\\
\bottomrule[0.7pt]
$\frac{5}{2}^{-}$ &
$49.6$&
$16320$&
$16544$&
$16318$&
$\frac{5}{2}^{-}$ &
$50.1$&
$16421$&
$16635$&
$16422$\\
$\frac{3}{2}^{-}$ &
$\begin{pmatrix}49.6\\16.6\\-94.5\end{pmatrix}$&
$\begin{pmatrix}16320\\16287\\16176\end{pmatrix}$&
$\begin{pmatrix}16544\\16511\\16400\end{pmatrix}$&
$\begin{pmatrix}16538\\16318\\16176\end{pmatrix}$&
$\frac{3}{2}^{-}$ &
$\begin{pmatrix}50.2\\16.1\\-94.4\end{pmatrix}$&
$\begin{pmatrix}16421\\16387\\16274\end{pmatrix}$&
$\begin{pmatrix}16636\\16601\\16489\end{pmatrix}$&
$\begin{pmatrix}16626\\16422\\16277\end{pmatrix}$\\
$\frac{1}{2}^{-}$ &
$\begin{pmatrix}72.2\\30.3\\-7.5\end{pmatrix}$&
$\begin{pmatrix}16343\\16301\\16263\end{pmatrix}$&
$\begin{pmatrix}16567\\16525\\16487\end{pmatrix}$&
$\begin{pmatrix}16574\\16523\\16315\end{pmatrix}$&
$\frac{1}{2}^{-}$ &
$\begin{pmatrix}74.0\\31.8\\-8.7\end{pmatrix}$&
$\begin{pmatrix}16445\\16403\\16362\end{pmatrix}$&
$\begin{pmatrix}16659\\16617\\16577\end{pmatrix}$&
$\begin{pmatrix}16663\\16611\\16418\end{pmatrix}$\\
\bottomrule[1.0pt]
\multicolumn{5}{l|}{$ccbb\bar{n}$}&\multicolumn{5}{l}{$ccbb\bar{s}$}\\
$J^P$&Eigenvalue&($B\Omega_{ccb}$)&($D\Omega_{bbc}$)&Mass&$J^P$&Eigenvalue&$(B_{s}\Omega_{ccb})$&$(D_{s}\Omega_{bbc})$&Mass\\
\bottomrule[0.7pt]
$\frac{5}{2}^{-}$ &
$42.1$&
$13358$&
$13199$&
$13244$&
$\frac{5}{2}^{-}$ &
$42.9$&
$13450$&
$13300$&
$13341$\\
$\frac{3}{2}^{-}$ &
$\begin{pmatrix}61.2\\24.3\\11.3\\-70.4\end{pmatrix}$&
$\begin{pmatrix}13377\\13340\\13327\\13246\end{pmatrix}$&
$\begin{pmatrix}13218\\13181\\13168\\13086\end{pmatrix}$&
$\begin{pmatrix}13383\\13231\\13221\\13100\end{pmatrix}$&
$\frac{3}{2}^{-}$ &
$\begin{pmatrix}61.7\\24.1\\12.0\\-72.6\end{pmatrix}$&
$\begin{pmatrix}13468\\13431\\13419\\13334\end{pmatrix}$&
$\begin{pmatrix}13319\\13281\\13269\\13185\end{pmatrix}$&
$\begin{pmatrix}13470\\13329\\13319\\13200\end{pmatrix}$\\
$\frac{1}{2}^{-}$ &
$\begin{pmatrix}90.3\\44.8\\-5.4\\-84.4\end{pmatrix}$&
$\begin{pmatrix}13406\\13361\\13311\\13232\end{pmatrix}$&
$\begin{pmatrix}13247\\13202\\13151\\13072\end{pmatrix}$&
$\begin{pmatrix}13414\\13242\\13212\\13086\end{pmatrix}$&
$\frac{1}{2}^{-}$ &
$\begin{pmatrix}91.7\\46.3\\-7.3\\-85.6\end{pmatrix}$&
$\begin{pmatrix}13498\\13453\\13399\\13321\end{pmatrix}$&
$\begin{pmatrix}13349\\13303\\13250\\13172\end{pmatrix}$&
$\begin{pmatrix}13502\\13343\\13307\\13187\end{pmatrix}$\\
\bottomrule[0.5pt]
\bottomrule[1.5pt]
\end{tabular}
\end{table*}

\begin{figure*}[t]
\begin{tabular}{cc}
\includegraphics[width=265pt]{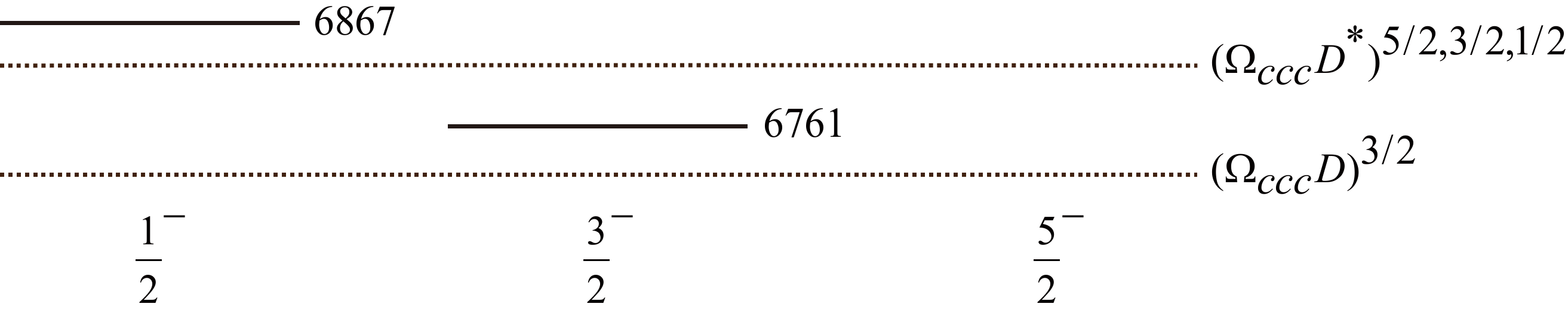}&
\includegraphics[width=265pt]{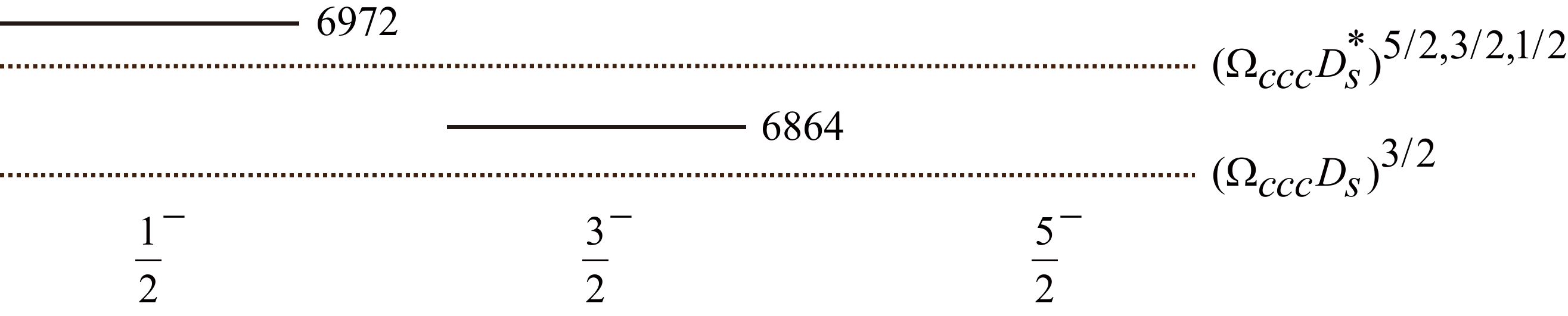}\\
(a) \begin{tabular}{c}  $cccc\bar{n}$ states\end{tabular} &(b)  $cccc\bar{s}$ states\\
\includegraphics[width=265pt]{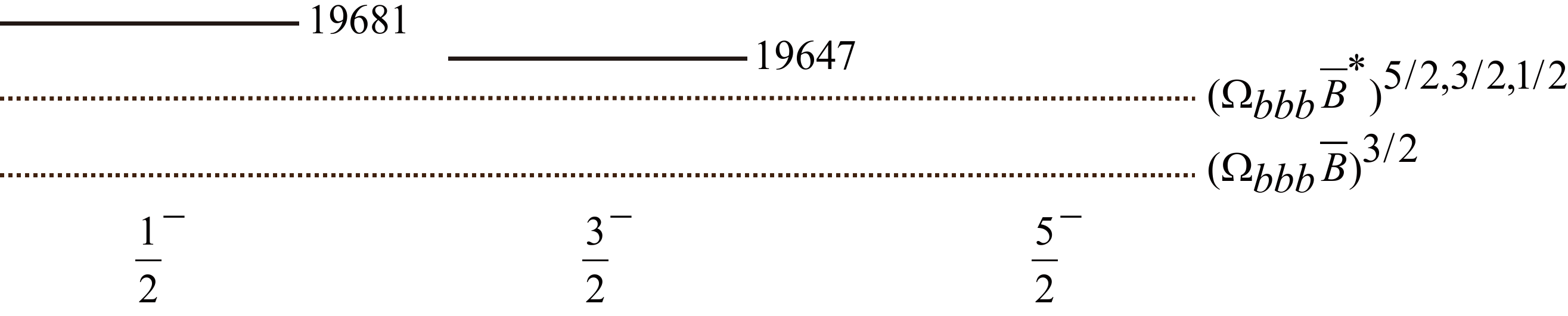}&
\includegraphics[width=265pt]{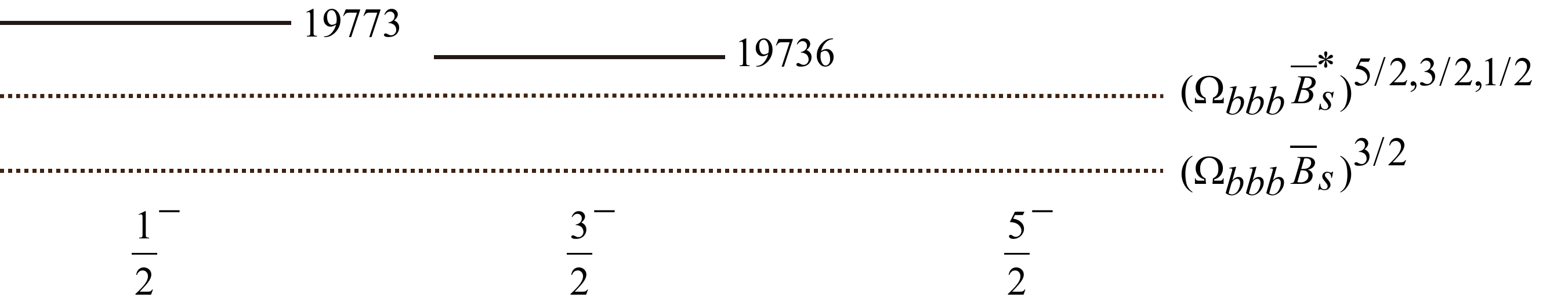}\\
(c) \begin{tabular}{c}  $bbbb\bar{n}$ states\end{tabular} &(d)  $bbbb\bar{s}$ states\\
\includegraphics[width=265pt]{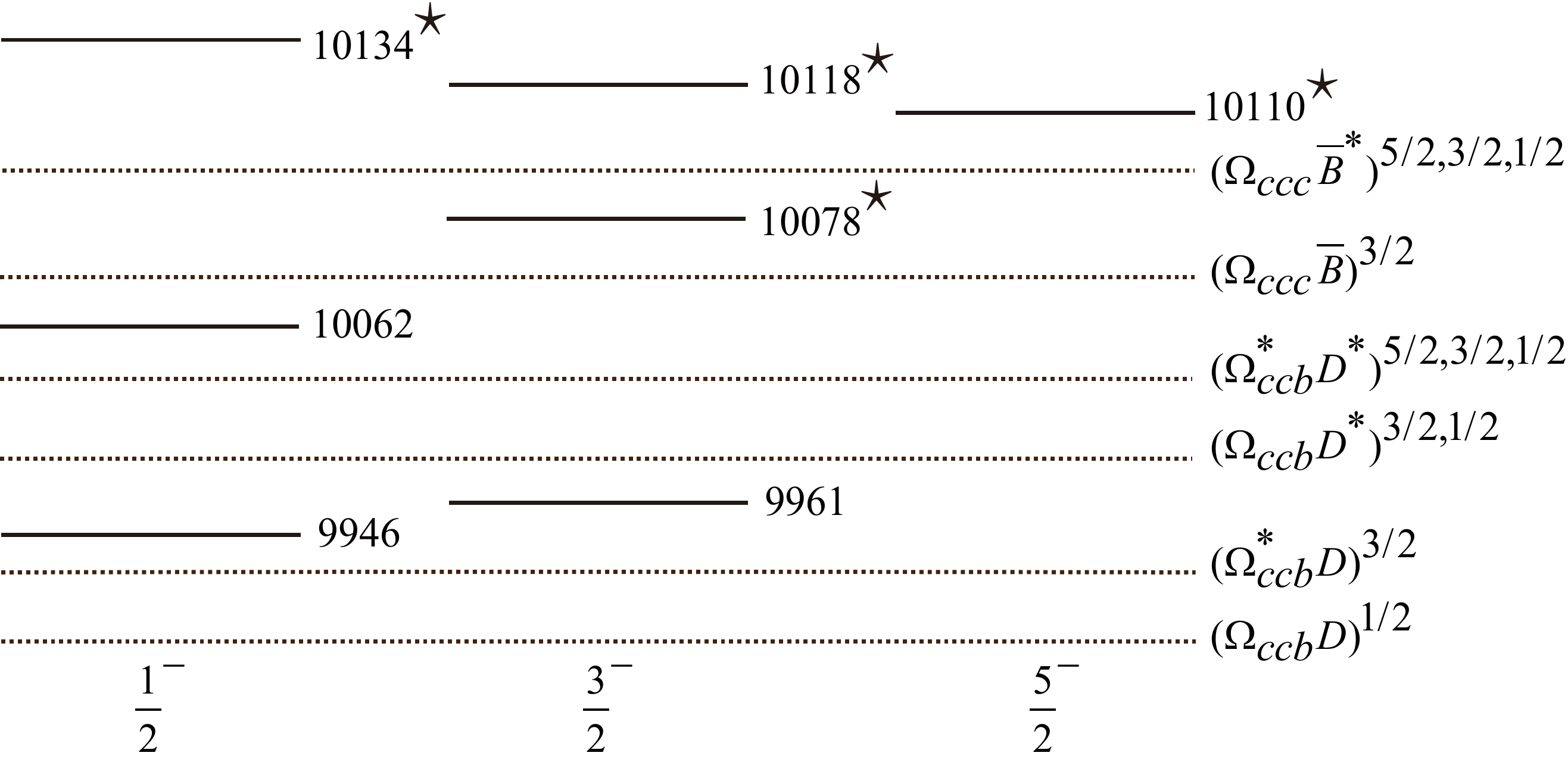}&
\includegraphics[width=265pt]{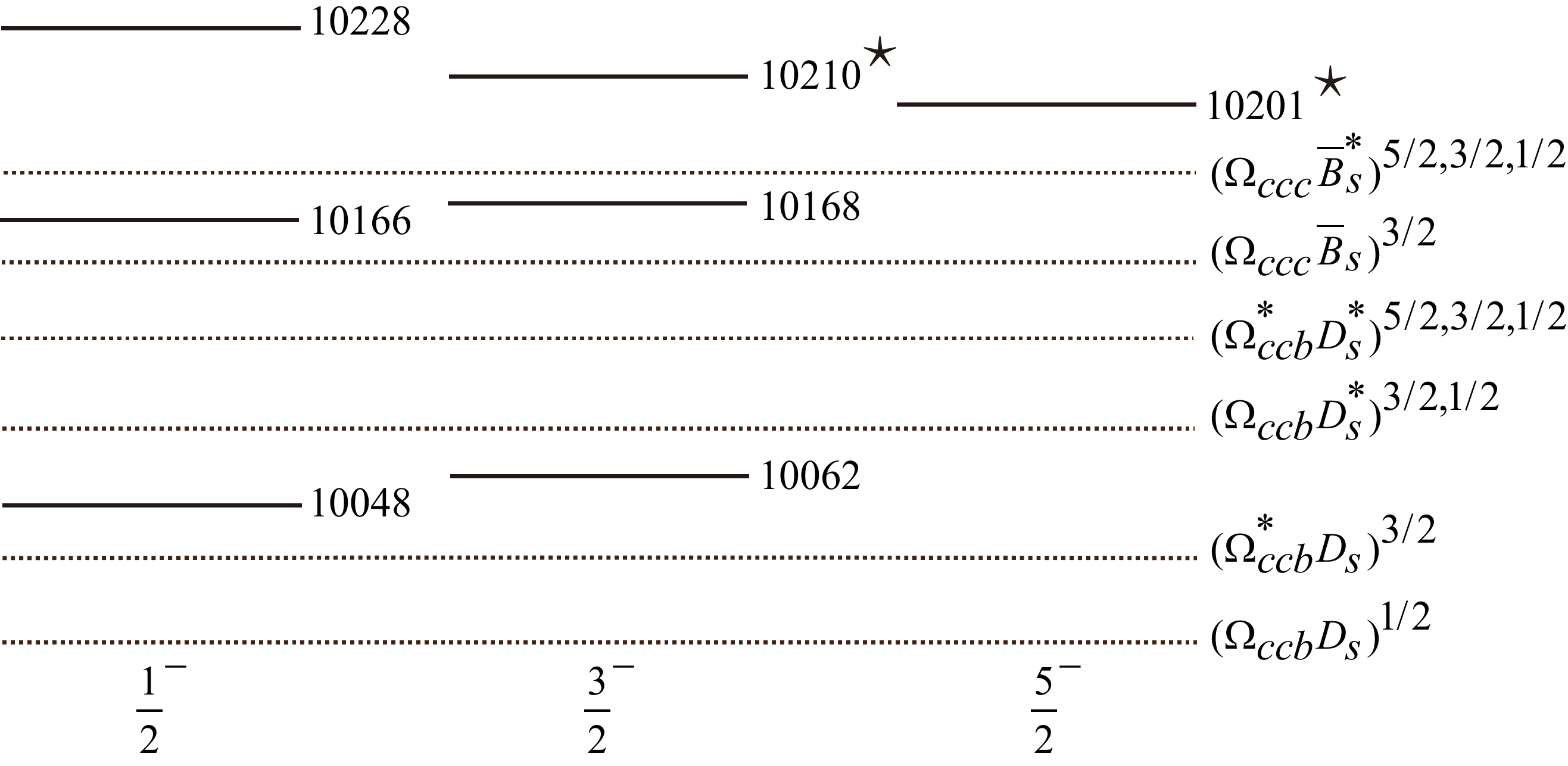}\\
(e) \begin{tabular}{c}  $cccb\bar{n}$ states\end{tabular} &(f)  $cccb\bar{s}$ states\\
\includegraphics[width=265pt]{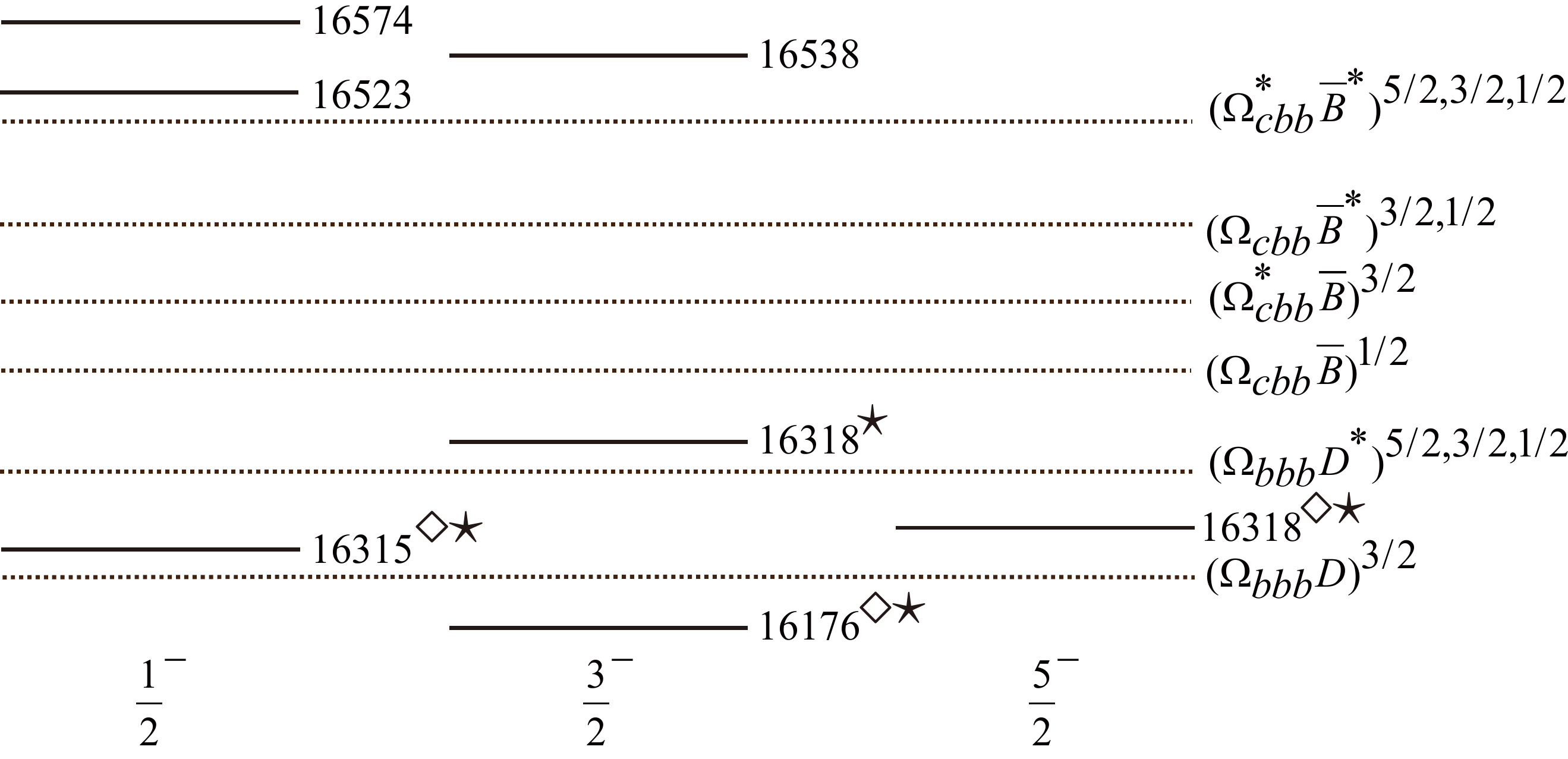}&
\includegraphics[width=265pt]{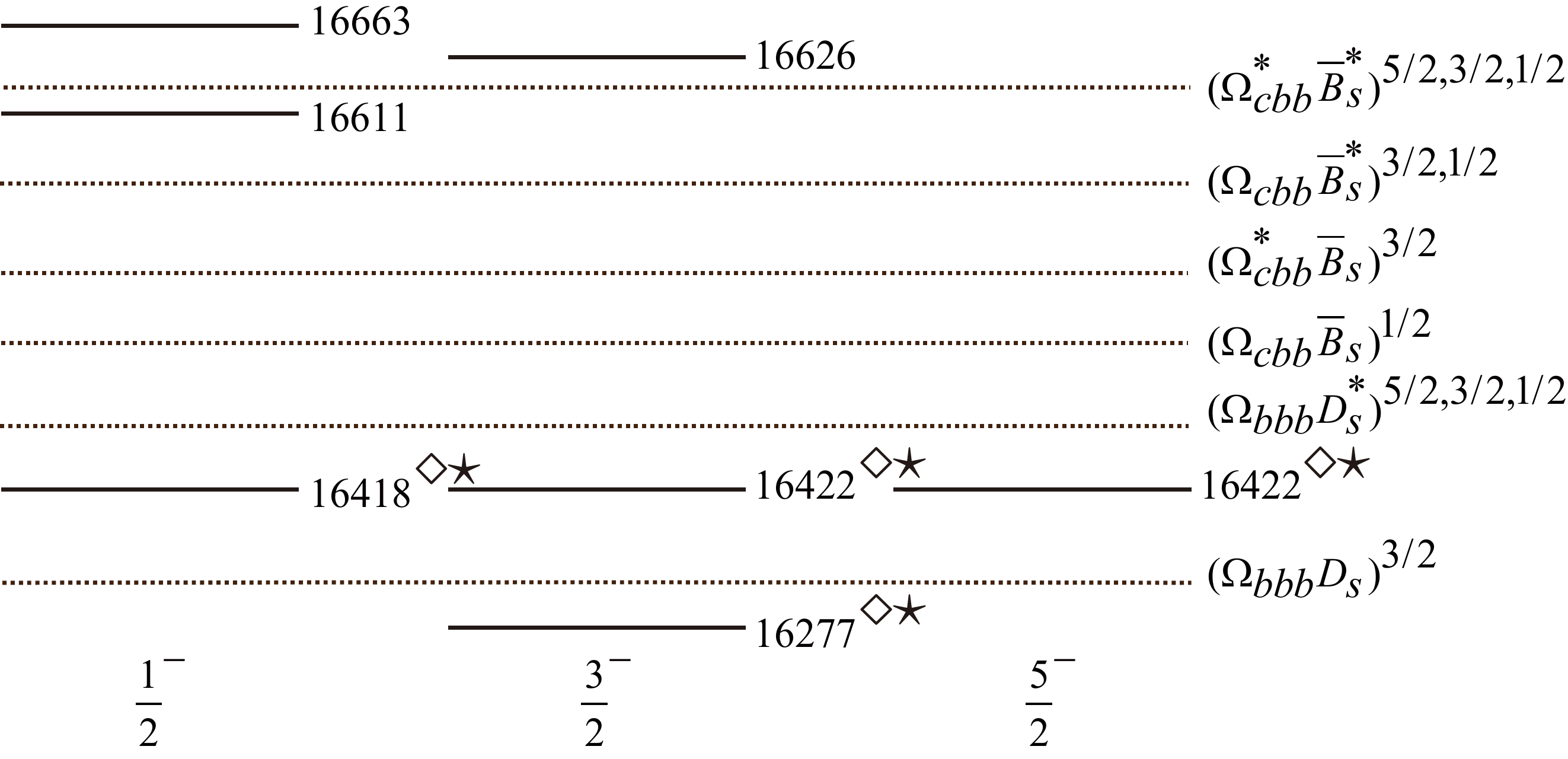}\\
(g) \begin{tabular}{c}  $bbbc\bar{n}$ states\end{tabular} &(h)  $bbbc\bar{s}$ states\\
\includegraphics[width=265pt]{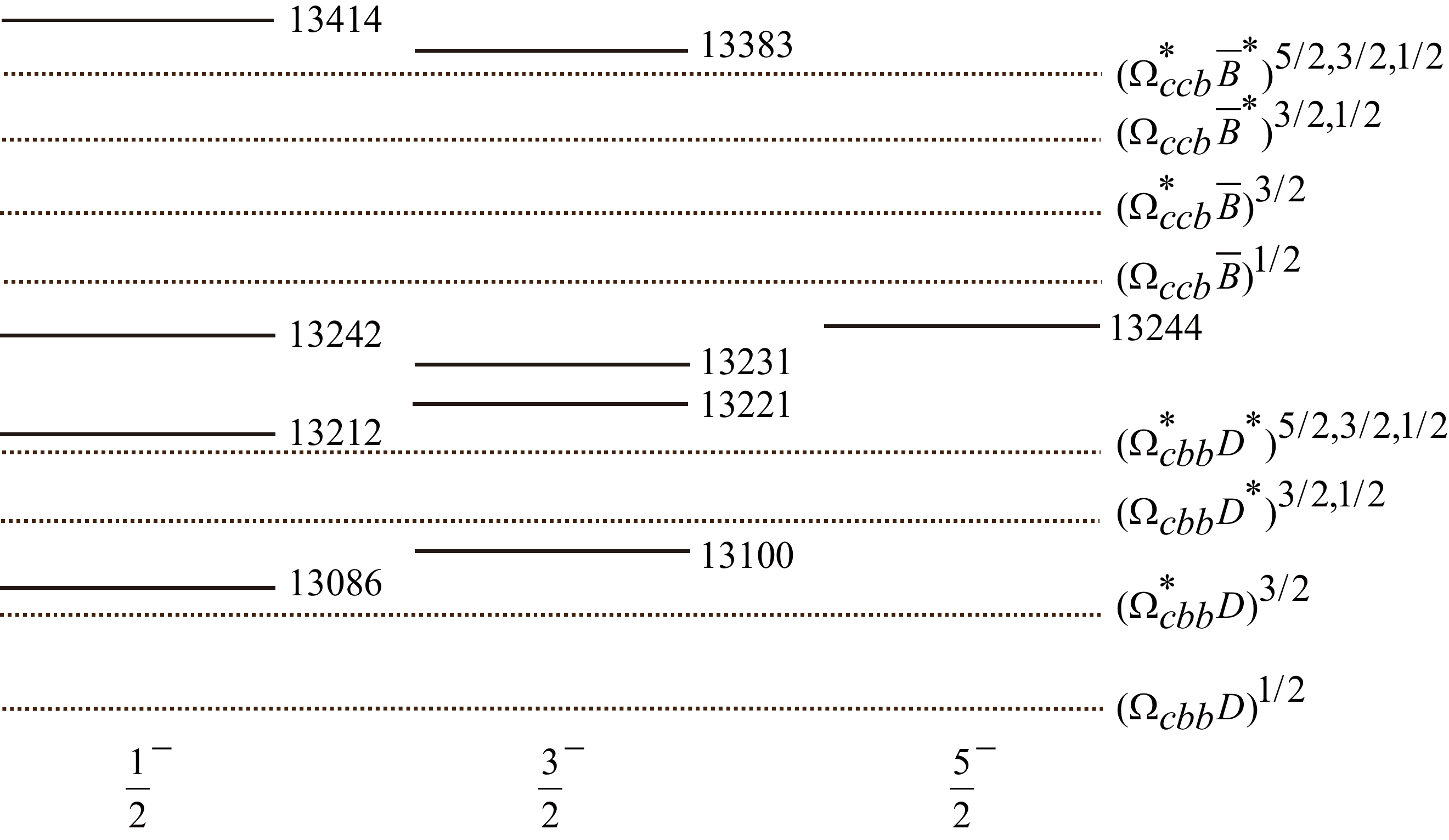}&
\includegraphics[width=265pt]{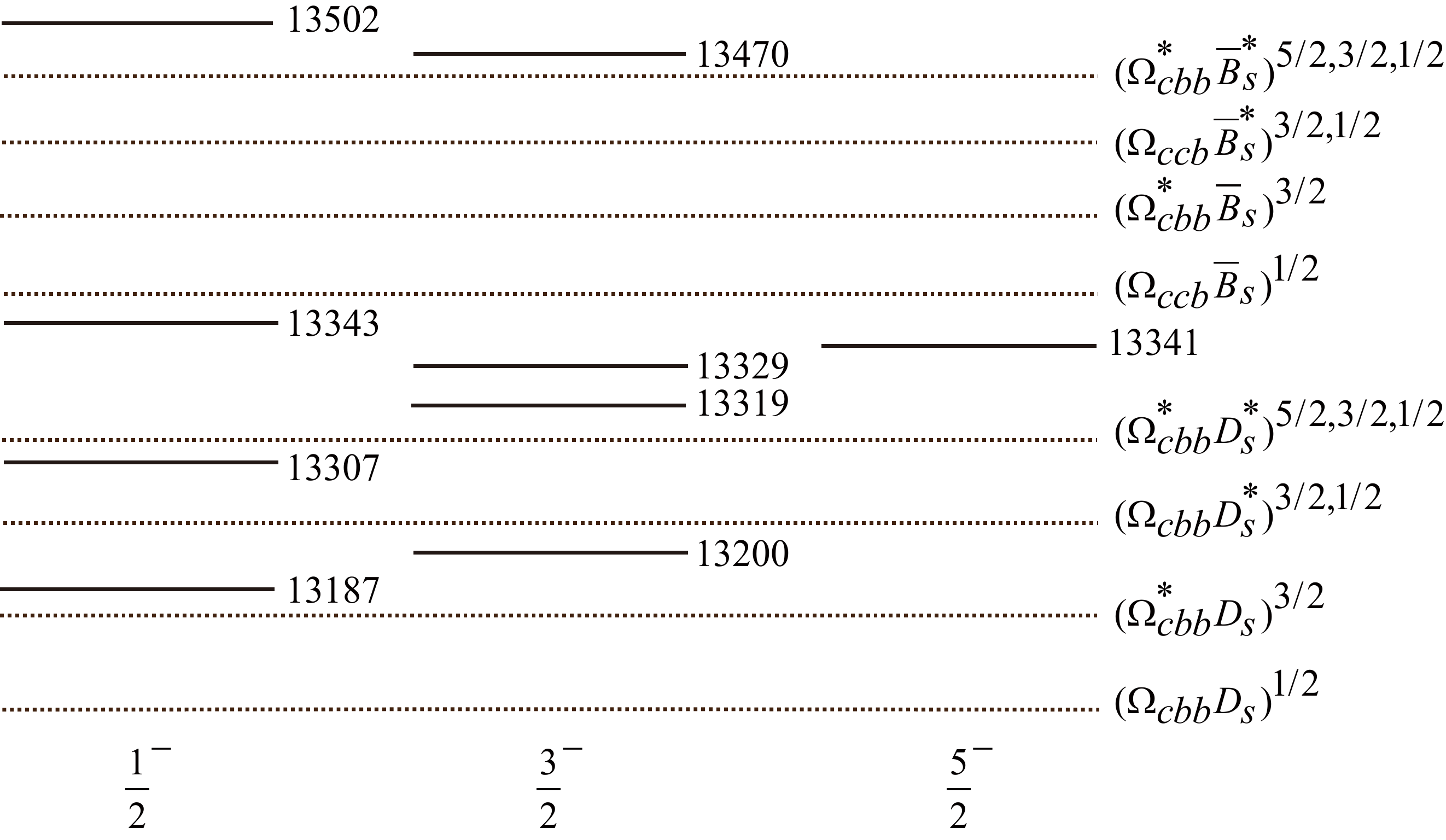}\\
(i) \begin{tabular}{c}  $ccbb\bar{n}$ states\end{tabular} &(j)  $ccbb\bar{s}$ states\\
\end{tabular}
\caption{
Relative positions (units: MeV) for the $cccc\bar{n}$, $cccc\bar{s}$, $bbbb\bar{n}$, $bbbb\bar{s}$, $cccb\bar{n}$, $cccb\bar{s}$, $bbbc\bar{n}$, $bbbc\bar{s}$, $ccbb\bar{n}$, and $ccbb\bar{s}$ pentaquark states labeled with solid lines.
The dotted lines denote various S-wave baryon-meson thresholds, and the superscripts of the labels, e.g. $(\Omega_{ccc}D^{*})^{5/2,3/2,1/2}$, represent the possible total angular momenta of the channels.
We mark the relatively stable pentaquarks, unable to decay into the S-wave baryon-meson states, with ``$\diamond$" after their masses. We mark the pentaquark whose wave function overlaps with that of one special baryon-meson state more than 90\% with
``$\star$" after their masses.
}\label{fig-QQQQq}
\end{figure*}

\begin{table*}[t]
\centering \caption{The eigenvectors of the $cccc\bar{n}$, $cccc\bar{s}$, $bbbb\bar{n}$, $bbbb\bar{s}$, $cccb\bar{n}$, $cccb\bar{s}$, $bbbc\bar{n}$, $bbbc\bar{s}$, $ccbb\bar{n}$, and $ccbb\bar{s}$ pentaquark subsystems. The masses are all in units of MeV. See the caption of Fig. \ref{fig-QQQQq} for meanings of ``$\diamond$" and ``$\star$".
}\label{eigenvector-QQQQq}
\renewcommand\arraystretch{1.15}
\begin{tabular}{cl|cc|cccc|l|cc|cccc}
\bottomrule[1.5pt]
\bottomrule[0.5pt]
&\multicolumn{1}{c}{}&\multicolumn{2}{c|}{$ccc\bigotimes c\bar{n}$}&&\multicolumn{2}{c|}{$ccc \bigotimes c\bar{s}$}&&\multicolumn{1}{c}{}&\multicolumn{2}{c|}{$bbb\bigotimes b\bar{n}$}&&\multicolumn{2}{c|}{$bbb \bigotimes b\bar{s}$}\\
$J^P$&Mass&$\Omega_{ccc}D^{*}$&\multicolumn{1}{c|}{$\Omega_{ccc}D$}&\multicolumn{1}{c|}{Mass}&$\Omega_{ccc} D_{s}^{*}$&\multicolumn{1}{c|}{$\Omega_{ccc} D_s$}
&&Mass&$\Omega_{bbb}B^{*}$&$\Omega_{bbb}B$&\multicolumn{1}{c|}{Mass}&$\Omega_{bbb} B_{s}^{*}$&\multicolumn{1}{c|}{$\Omega_{bbb} B_s$}\\
\Xcline{1-7}{0.7pt}\Xcline{9-14}{0.7pt}
$\frac{3}{2}^{-}$&6761&0.456&-0.354&\multicolumn{1}{c|}{6864}&0.456&\multicolumn{1}{c|}{-0.354}&&19647&0.456&-0.354&\multicolumn{1}{c|}{19736}&0.456&\multicolumn{1}{c|}{-0.354}\\
$\frac{1}{2}^{-}$&6867&-0.577&&\multicolumn{1}{c|}{6972}&0.577&\multicolumn{1}{c|}{}&&19681&-0.577&&\multicolumn{1}{c|}{19773}&0.577&\multicolumn{1}{c|}{}\\
\bottomrule[0.5pt]
\multicolumn{2}{l|}{}&\multicolumn{2}{c}{$ccc\bigotimes b\bar{n}$}&\multicolumn{4}{c|}{$ccb\bigotimes c\bar{n}$}&\multicolumn{1}{l}{}&\multicolumn{2}{c}{$ccc\bigotimes b\bar{s}$}&\multicolumn{4}{c}{$ccb\bigotimes c\bar{s}$}\\
$J^P$&Mass&$\Omega_{ccc} B^{*}$&$\Omega_{ccc} B$
&$\Omega_{ccb}^{*}D^{*}$&$\Omega_{ccb}^{*}D$&$\Omega_{ccb}D^{*}$&$\Omega_{ccb}D$&
Mass&$\Omega_{ccc} B_{s}^{*}$&$\Omega_{ccc} B_{s}$
&$\Omega_{ccb}^{*}D_{s}^{*}$&$\Omega_{ccb}^{*}D_{s}$&$\Omega_{ccb}D_{s}^{*}$&$\Omega_{ccb}D_{s}$\\
\bottomrule[0.7pt]
$\frac{5}{2}^{-}$&$10110\star$&1.000&&0.333&&&&$10201\star$&1.000&&0.333&&&\\
$\frac{3}{2}^{-}$&$10118\star$&0.950&0.188&0.217&0.140&-0.255&&$10210\star$&0.939&0.197&0.234&0.127&-0.258\\
&$10078\star$&-0.261&0.917&0.333&-0.019&0.174&&10168&-0.289&0.895&0.345&-0.043&0.169\\
&$9961$&0.172&0.352&0.372&-0.450&-0.230&&10062&0.184&0.399&0.350&-0.452&-0.231\\
$\frac{1}{2}^{-}$
&$10134\star$&0.914&&-0.333&&-0.201&-0.121&10228&0.891&&-0.365&&-0.188&-0.106\\
&10062&-0.246&&0.498&&-0.415&-0.064&10166&-0.290&&0.477&&-0.430&-0.065\\
&9946&0.323&&0.220&&0.397&-0.451&10048&0.349&&0.217&&0.387&-0.455\\
\bottomrule[0.5pt]
\multicolumn{2}{l|}{}&\multicolumn{2}{c}{$bbb\bigotimes c\bar{n}$}&\multicolumn{4}{c|}{$bbc\bigotimes b\bar{n}$}&\multicolumn{1}{l}{}&\multicolumn{2}{c}{$bbb\bigotimes c\bar{s}$}&\multicolumn{4}{c}{$bbc\bigotimes b\bar{s}$}\\
$J^P$&Mass&$\Omega_{bbb} D^{*}$&$\Omega_{bbb} D$
&$\Omega_{bbc}^{*}B^{*}$&$\Omega_{bbc}^{*}B$&$\Omega_{bbc}B^{*}$&$\Omega_{bbc}B$&
Mass&$\Omega_{bbb} D_{s}^{*}$&$\Omega_{bbb} D_{s}$
&$\Omega_{bbc}^{*}B_{s}^{*}$&$\Omega_{bbc}^{*}B_{s}$&$\Omega_{bbc}B_{s}^{*}$&$\Omega_{bbc}B_{s}$\\
\bottomrule[0.7pt]
$\frac{5}{2}^{-}$
&$16318\star\diamond$&1.000&&0.333&&&&$16422\star\diamond$&1.000&&0.333&&&\\
$\frac{3}{2}^{-}$
&16538&-0.004&-0.052&0.508&-0.375&-0.213&&16626&-0.009&-0.056&0.509&-0.373&-0.213\\
&$16318\star$&0.999&-0.001&0.053&0.217&-0.139&&$16422\star\diamond$&0.999&-0.001&0.051&0.218&-0.140\\
&$16176\star\diamond$&0.001&0.999&0.189&0.187&0.204&&$16277\star\diamond$&0.001&0.998&0.187&0.188&0.204\\
$\frac{1}{2}^{-}$
&16574&-0.085&&-0.634&&0.135&0.146&16663&-0.092&&-0.634&&0.139&0.141\\
&16523&-0.074&&0.045&&0.580&-0.323&16611&-0.086&&0.052&&0.580&-0.321\\
&$16315\star\diamond$&0.994&&0.054&&0.127&-0.130&$16418\star\diamond$&0.992&&0.049&&0.121&0.315\\
\bottomrule[0.5pt]
\multicolumn{2}{l|}{}&\multicolumn{4}{c}{$bbc\bigotimes c\bar{n}$}&\multicolumn{4}{c|}{$ccb\bigotimes b\bar{n}$}\\
$J^P$&Mass&
$\Omega^{*}_{bbc}D^{*}$&\multicolumn{1}{c}{$\Omega^{*}_{bbc}D$}&$\Omega_{bbc}D^{*}$&\multicolumn{1}{c|}{$\Omega_{bbc}D$}&
$\Omega^{*}_{ccb}B^{*}$&\multicolumn{1}{c}{$\Omega^{*}_{ccb}B$}&\multicolumn{1}{c}{$\Omega_{ccb}B^{*}$}&\multicolumn{1}{c|}{$\Omega_{ccb}B$}\\
\Xcline{1-10}{0.7pt}
$\frac52^{-}$&13244&-0.577&\multicolumn{1}{c}{}&&\multicolumn{1}{c|}{}&-0.577&\multicolumn{1}{c}{}&\multicolumn{1}{c}{}&\multicolumn{1}{c|}{}\\
$\frac32^{-}$
&13383&-0.018&\multicolumn{1}{c}{-0.052}&-0.013&\multicolumn{1}{c|}{}&0.566&\multicolumn{1}{c}{-0.372}&\multicolumn{1}{c}{0.452}&\multicolumn{1}{c|}{}\\
&13231&0.621&\multicolumn{1}{c}{0.033}&-0.072&\multicolumn{1}{c|}{}&-0.077&\multicolumn{1}{c}{-0.454}&\multicolumn{1}{c}{-0.251}&\multicolumn{1}{c|}{}\\
&13221&-0.155&\multicolumn{1}{c}{-0.167}&-0.586&\multicolumn{1}{c|}{}&-0.319&\multicolumn{1}{c}{-0.040}&\multicolumn{1}{c}{0.410}&\multicolumn{1}{c|}{}\\
&13100&0.250&\multicolumn{1}{c}{-0.620}&0.310&\multicolumn{1}{c|}{}&0.209&\multicolumn{1}{c}{0.265}&\multicolumn{1}{c}{-0.096}&\multicolumn{1}{c|}{}\\
$\frac12^{-}$
&13414&-0.099&\multicolumn{1}{c}{}&0.053&\multicolumn{1}{c|}{0.024}&-0.710&\multicolumn{1}{c}{}&\multicolumn{1}{c}{-0.222}&\multicolumn{1}{c|}{-0.317}\\
&13242&0.641&\multicolumn{1}{c}{}&0.220&\multicolumn{1}{c|}{-0.107}&-0.124&\multicolumn{1}{c}{}&\multicolumn{1}{c}{0.385}&\multicolumn{1}{c|}{0.180}\\
&13212&0.197&\multicolumn{1}{c}{}&-0.535&\multicolumn{1}{c|}{0.052}&-0.078&\multicolumn{1}{c}{}&\multicolumn{1}{c}{-0.271}&\multicolumn{1}{c|}{0.510}\\
&13086&0.310&\multicolumn{1}{c}{}&0.155&\multicolumn{1}{c|}{0.634}&0.173&\multicolumn{1}{c}{}&\multicolumn{1}{c}{-0.301}&\multicolumn{1}{c|}{-0.155}\\
\Xcline{1-10}{0.5pt}
\multicolumn{2}{l|}{}&\multicolumn{4}{c}{$bbc\bigotimes c\bar{s}$}&\multicolumn{4}{c|}{$ccb\bigotimes b\bar{s}$}\\
$J^P$&Mass&
$\Omega^{*}_{bbc}D_{s}^{*}$&\multicolumn{1}{c}{$\Omega^{*}_{bbc}D_{s}$}&$\Omega_{bbc}D_{s}^{*}$&\multicolumn{1}{c|}{$\Omega_{bbc}D_{s}$}&
$\Omega^{*}_{ccb}B_{s}^{*}$&\multicolumn{1}{c}{$\Omega^{*}_{ccb}B_{s}$}&\multicolumn{1}{c}{$\Omega_{ccb}B_{s}^{*}$}&\multicolumn{1}{c|}{$\Omega_{ccb}B_{s}$}\\
\Xcline{1-10}{0.7pt}
$\frac52^{-}$&13341&-0.577&\multicolumn{1}{c}{}&&\multicolumn{1}{c|}{}&-0.577&\multicolumn{1}{c}{}&\multicolumn{1}{c}{}&\multicolumn{1}{c|}{}\\
$\frac32^{-}$
&13470&-0.025&\multicolumn{1}{c}{-0.056}&-0.013&\multicolumn{1}{c|}{}&0.569&\multicolumn{1}{c}{-0.366}&\multicolumn{1}{c}{0.453}&\multicolumn{1}{c|}{}\\
&13329&0.647&\multicolumn{1}{c}{0.041}&0.051&\multicolumn{1}{c|}{}&-0.122&\multicolumn{1}{c}{-0.448}&\multicolumn{1}{c}{-0.167}&\multicolumn{1}{c|}{}\\
&13319&-0.041&\multicolumn{1}{c}{-0.142}&-0.596&\multicolumn{1}{c|}{}&-0.292&\multicolumn{1}{c}{0.052}&\multicolumn{1}{c}{0.448}&\multicolumn{1}{c|}{}\\
&13200&-0.226&\multicolumn{1}{c}{0.626}&-0.293&\multicolumn{1}{c|}{}&0.218&\multicolumn{1}{c}{0.282}&\multicolumn{1}{c}{-0.104}&\multicolumn{1}{c|}{}\\
$\frac12^{-}$
&13502&-0.109&\multicolumn{1}{c}{}&0.057&\multicolumn{1}{c|}{0.025}&-0.713&\multicolumn{1}{c}{}&\multicolumn{1}{c}{-0.220}&\multicolumn{1}{c|}{-0.308}\\
&13343&0.648&\multicolumn{1}{c}{}&0.232&\multicolumn{1}{c|}{-0.085}&-0.108&\multicolumn{1}{c}{}&\multicolumn{1}{c}{0.380}&\multicolumn{1}{c|}{0.172}\\
&13307&0.201&\multicolumn{1}{c}{}&-0.531&\multicolumn{1}{c|}{0.055}&-0.069&\multicolumn{1}{c}{}&\multicolumn{1}{c}{-0.265}&\multicolumn{1}{c|}{0.515}\\
&13187&0.286&\multicolumn{1}{c}{}&0.150&\multicolumn{1}{c|}{0.637}&-0.177&\multicolumn{1}{c}{}&\multicolumn{1}{c}{0.313}&\multicolumn{1}{c|}{0.164}\\
\bottomrule[0.5pt]
\bottomrule[1.5pt]
\end{tabular}
\end{table*}

\begin{table*}[t]
\centering \caption{
The values of $k\cdot |c_{i}|^{2}$ for the the $cccc\bar{n}$, $cccc\bar{s}$, $bbbb\bar{n}$, $bbbb\bar{s}$, $cccb\bar{n}$, $cccb\bar{s}$, $bbbc\bar{n}$, $bbbc\bar{s}$, $ccbb\bar{n}$, and $ccbb\bar{s}$ pentaquark states. The masses are all in units of MeV.
The decay channel is marked with ``$\times$'' if kinetically forbidden.
See the caption of Fig. \ref{fig-QQQQq} for meanings of ``$\diamond$" and ``$\star$".
One can roughly estimate the relative decay widths between different decay processes of different initial pentaquark states with this table if neglecting the $\gamma_i$ differences.
}\label{value-QQQQq}
\renewcommand\arraystretch{1.15}
\begin{tabular}{cl|cc|cccc|l|cc|cccc}
\bottomrule[1.5pt]
\bottomrule[0.5pt]
&\multicolumn{1}{c}{}&\multicolumn{2}{c|}{$ccc\bigotimes c\bar{n}$}&&\multicolumn{2}{c|}{$ccc \bigotimes c\bar{s}$}&&\multicolumn{1}{c}{}&\multicolumn{2}{c|}{$bbb\bigotimes b\bar{n}$}&&\multicolumn{2}{c|}{$bbb \bigotimes b\bar{s}$}\\
$J^P$&Mass&$\Omega_{ccc}D^{*}$&\multicolumn{1}{c|}{$\Omega_{ccc}D$}&\multicolumn{1}{c|}{Mass}&$\Omega_{ccc} D_{s}^{*}$&\multicolumn{1}{c|}{$\Omega_{ccc} D_s$}
&&Mass&$\Omega_{bbb}B^{*}$&$\Omega_{bbb}B$&\multicolumn{1}{c|}{Mass}&$\Omega_{bbb} B_{s}^{*}$&\multicolumn{1}{c|}{$\Omega_{bbb} B_s$}\\
\Xcline{1-7}{0.7pt}\Xcline{9-14}{0.7pt}
$\frac{3}{2}^{-}$&6761&$\times$&67&\multicolumn{1}{c|}{6864}&$\times$&\multicolumn{1}{c|}{70}&&19647&66&84&\multicolumn{1}{c|}{19736}&62&\multicolumn{1}{c|}{86}\\
$\frac{1}{2}^{-}$&6867&151&&\multicolumn{1}{c|}{6972}&156&\multicolumn{1}{c|}{}&&19681&201&&\multicolumn{1}{c|}{19773}&204&\multicolumn{1}{c|}{}\\
\bottomrule[0.5pt]
\multicolumn{2}{l|}{}&\multicolumn{2}{c}{$ccc\bigotimes b\bar{n}$}&\multicolumn{4}{c|}{$ccb\bigotimes c\bar{n}$}&\multicolumn{1}{l}{}&\multicolumn{2}{c}{$ccc\bigotimes b\bar{s}$}&\multicolumn{4}{c}{$ccb\bigotimes c\bar{s}$}\\
$J^P$&Mass&$\Omega_{ccc} B^{*}$&$\Omega_{ccc} B$
&$\Omega_{ccb}^{*}D^{*}$&$\Omega_{ccb}^{*}D$&$\Omega_{ccb}D^{*}$&$\Omega_{ccb}D$&
Mass&$\Omega_{ccc} B_{s}^{*}$&$\Omega_{ccc} B_{s}$
&$\Omega_{ccb}^{*}D_{s}^{*}$&$\Omega_{ccb}^{*}D_{s}$&$\Omega_{ccb}D_{s}^{*}$&$\Omega_{ccb}D_{s}$\\
\bottomrule[0.7pt]
$\frac{5}{2}^{-}$
&$10110\star$&$\times$&&56&&&&$10201\star$&13&&53&&&\\
$\frac{3}{2}^{-}$
&$10118\star$&173&18&25&17&40&&$10210\star$&184&21&28&14&40\\
&$10078\star$&$\times$&217&43&0.3&15&&10168&$\times$&223&40&1&13\\
&$9961$&$\times$&$\times$&$\times$&94&$\times$&&10062&$\times$&$\times$&$\times$&98&$\times$\\
$\frac{1}{2}^{-}$
&$10134\star$&291&&64&&27&14&10228&296&&75&&23&11\\
&10062&$\times$&&76&&77&3&10166&$\times$&&75&&86&3\\
&9946&$\times$&&$\times$&&$\times$&104&10048&$\times$&&$\times$&&$\times$&111\\
\bottomrule[0.5pt]
\multicolumn{2}{l|}{}&\multicolumn{2}{c}{$bbb\bigotimes c\bar{n}$}&\multicolumn{4}{c|}{$bbc\bigotimes b\bar{n}$}&\multicolumn{1}{l}{}&\multicolumn{2}{c}{$bbb\bigotimes c\bar{s}$}&\multicolumn{4}{c}{$bbc\bigotimes b\bar{s}$}\\
$J^P$&Mass&$\Omega_{bbb} D^{*}$&$\Omega_{bbb} D$
&$\Omega_{bbc}^{*}B^{*}$&$\Omega_{bbc}^{*}B$&$\Omega_{bbc}B^{*}$&$\Omega_{bbc}B$&
Mass&$\Omega_{bbb} D_{s}^{*}$&$\Omega_{bbb} D_{s}$
&$\Omega_{bbc}^{*}B_{s}^{*}$&$\Omega_{bbc}^{*}B_{s}$&$\Omega_{bbc}B_{s}^{*}$&$\Omega_{bbc}B_{s}$\\
\bottomrule[0.7pt]
$\frac{5}{2}^{-}$
&$16318\star\diamond$&$\times$&&$\times$&&&&$16422\star\diamond$&$\times$&&$\times$&&&\\
$\frac{3}{2}^{-}$
&16538&0.02&3&89&94&27&&16626&0.1&4&82&94&26\\
&$16318\star$&$\times$&0.001&$\times$&$\times$&$\times$&&$16422\star\diamond$&$\times$&$\times$&$\times$&$\times$&$\times$\\
&$16176\star\diamond$&$\times$&$\times$&$\times$&$\times$&$\times$&&$16277\star\diamond$&$\times$&$\times$&$\times$&$\times$&$\times$\\
$\frac{1}{2}^{-}$
&16574&7&&248&&14&21&16663&8&&247&&15&20\\
&16523&5&&0.2&&164&79&16611&6&&$\times$&&160&78\\
&$16315\star\diamond$&$\times$&&$\times$&&$\times$&$\times$&$16418\star\diamond$&$\times$&&$\times$&&$\times$&$\times$\\
\bottomrule[0.5pt]
\multicolumn{2}{l|}{}&\multicolumn{4}{c}{$bbc\bigotimes c\bar{n}$}&\multicolumn{4}{c|}{$ccb\bigotimes b\bar{n}$}\\
$J^P$&Mass&
$\Omega^{*}_{bbc}D^{*}$&\multicolumn{1}{c}{$\Omega^{*}_{bbc}D$}&$\Omega_{bbc}D^{*}$&\multicolumn{1}{c|}{$\Omega_{bbc}D$}&
$\Omega^{*}_{ccb}B^{*}$&\multicolumn{1}{c}{$\Omega^{*}_{ccb}B$}&\multicolumn{1}{c}{$\Omega_{ccb}B^{*}$}&\multicolumn{1}{c|}{$\Omega_{ccb}B$}\\
\Xcline{1-10}{0.7pt}
$\frac52^{-}$&13244&119&\multicolumn{1}{c}{}&&\multicolumn{1}{c|}{}&$\times$&\multicolumn{1}{c}{}&\multicolumn{1}{c}{}&\multicolumn{1}{c|}{}\\
$\frac32^{-}$
&13383&0.3&\multicolumn{1}{c}{3}&0.2&\multicolumn{1}{c|}{}&154&\multicolumn{1}{c}{100}&\multicolumn{1}{c}{135}&\multicolumn{1}{c|}{}\\
&13231&111&\multicolumn{1}{c}{0.8}&2&\multicolumn{1}{c|}{}&$\times$&\multicolumn{1}{c}{$\times$}&\multicolumn{1}{c}{$\times$}&\multicolumn{1}{c|}{}\\
&13221&5&\multicolumn{1}{c}{20}&135&\multicolumn{1}{c|}{}&$\times$&\multicolumn{1}{c}{$\times$}&\multicolumn{1}{c}{$\times$}&\multicolumn{1}{c|}{}\\
&13100&$\times$&\multicolumn{1}{c}{127}&$\times$&\multicolumn{1}{c|}{}&$\times$&\multicolumn{1}{c}{$\times$}&\multicolumn{1}{c}{$\times$}&\multicolumn{1}{c|}{}\\
$\frac12^{-}$
&13414&8&\multicolumn{1}{c}{}&3&\multicolumn{1}{c|}{0.7}&332&\multicolumn{1}{c}{}&\multicolumn{1}{c}{39}&\multicolumn{1}{c|}{96}\\
&13242&142&\multicolumn{1}{c}{}&23&\multicolumn{1}{c|}{9}&$\times$&\multicolumn{1}{c}{}&\multicolumn{1}{c}{$\times$}&\multicolumn{1}{c|}{$\times$}\\
&13212&5&\multicolumn{1}{c}{}&101&\multicolumn{1}{c|}{2}&$\times$&\multicolumn{1}{c}{}&\multicolumn{1}{c}{$\times$}&\multicolumn{1}{c|}{$\times$}\\
&13086&$\times$&\multicolumn{1}{c}{}&$\times$&\multicolumn{1}{c|}{164}&$\times$&\multicolumn{1}{c}{}&\multicolumn{1}{c}{$\times$}&\multicolumn{1}{c|}{$\times$}\\
\Xcline{1-10}{0.5pt}
\multicolumn{2}{l|}{}&\multicolumn{4}{c}{$bbc\bigotimes c\bar{s}$}&\multicolumn{4}{c|}{$ccb\bigotimes b\bar{s}$}\\
$J^P$&Mass&
$\Omega^{*}_{bbc}D_{s}^{*}$&\multicolumn{1}{c}{$\Omega^{*}_{bbc}D_{s}$}&$\Omega_{bbc}D_{s}^{*}$&\multicolumn{1}{c|}{$\Omega_{bbc}D_{s}$}&
$\Omega^{*}_{ccb}B_{s}^{*}$&\multicolumn{1}{c}{$\Omega^{*}_{ccb}B_{s}$}&\multicolumn{1}{c}{$\Omega_{ccb}B_{s}^{*}$}&\multicolumn{1}{c|}{$\Omega_{ccb}B_{s}$}\\
\Xcline{1-10}{0.7pt}
$\frac52^{-}$&13341&113&\multicolumn{1}{c}{}&&\multicolumn{1}{c|}{}&$\times$&\multicolumn{1}{c}{}&\multicolumn{1}{c}{}&\multicolumn{1}{c|}{}\\
$\frac32^{-}$
&13470&8&\multicolumn{1}{c}{3}&0.7&\multicolumn{1}{c|}{}&332&\multicolumn{1}{c}{39}&\multicolumn{1}{c}{96}&\multicolumn{1}{c|}{}\\
&13329&142&\multicolumn{1}{c}{23}&9&\multicolumn{1}{c|}{}&$\times$&\multicolumn{1}{c}{$\times$}&\multicolumn{1}{c}{$\times$}&\multicolumn{1}{c|}{}\\
&13319&5&\multicolumn{1}{c}{101}&2&\multicolumn{1}{c|}{}&$\times$&\multicolumn{1}{c}{$\times$}&\multicolumn{1}{c}{$\times$}&\multicolumn{1}{c|}{}\\
&13200&$\times$&\multicolumn{1}{c}{$\times$}&164&\multicolumn{1}{c|}{}&$\times$&\multicolumn{1}{c}{$\times$}&\multicolumn{1}{c}{$\times$}&\multicolumn{1}{c|}{}\\
$\frac12^{-}$
&13502&10&\multicolumn{1}{c}{}&3&\multicolumn{1}{c|}{0.7}&328&\multicolumn{1}{c}{}&\multicolumn{1}{c}{38}&\multicolumn{1}{c|}{92}\\
&13343&148&\multicolumn{1}{c}{}&26&\multicolumn{1}{c|}{6}&$\times$&\multicolumn{1}{c}{}&\multicolumn{1}{c}{$\times$}&\multicolumn{1}{c|}{$\times$}\\
&13307&$\times$&\multicolumn{1}{c}{}&92&\multicolumn{1}{c|}{2}&$\times$&\multicolumn{1}{c}{}&\multicolumn{1}{c}{$\times$}&\multicolumn{1}{c|}{$\times$}\\
&13187&$\times$&\multicolumn{1}{c}{}&$\times$&\multicolumn{1}{c|}{173}&$\times$&\multicolumn{1}{c}{}&\multicolumn{1}{c}{$\times$}&\multicolumn{1}{c|}{$\times$}\\
\bottomrule[0.5pt]
\bottomrule[1.5pt]
\end{tabular}
\end{table*}

\subsubsection{The $cccc\bar{q}$ and $bbbb\bar{q}$ pentaquark states}
Here, we first discuss the $cccc\bar{q}$ and $bbbb\bar{q}$ pentaquark subsystems.
Because of the symmetrical constraint from Pauli principle, i.e., fully antisymmetric among the first four charm quarks, the ground $J^P=5/2^{-}$ pentaquark state with $cccc\bar{q}$ and $bbbb\bar{q}$ can not exist.
We only find two ground states: a $J^{P}=3/2^{-}$ state and a $J^{P}=1/2^{-}$ state for these subsystems.

Diagonalizing the Hamiltonians in Table \ref{nnnsQ} with the corresponding parameters in Table \ref{parameter2}, we can obtain the corresponding mass spectra for $cccc\bar{q}$ and $bbbb\bar{q}$ pentaquark subsystems and present them in Table \ref{mass-QQQQq}.
For the reference mass scheme, the only combination of meson-baryon reference systems are ($\Omega_{ccc}$)+($D$), ($\Omega_{ccc}$)+($D_{s}$), ($\Omega_{bbb}$)+($\bar{B}$), and ($\Omega_{bbb}$)+($\bar{B}_{s}$) for the $cccc\bar{n}$, $cccc\bar{s}$, $bbbb\bar{n}$, and $bbbb\bar{s}$ subsystems, respectively.

The modified CMI model scheme takes the chromoelectric interaction explicitly compared to the reference mass scheme, and therefore we use the results in this scheme for the following analysis.
Based on the results calculated from the modified CMI model scheme, we plot the mass spectra of the $cccc\bar{n}$, $cccc\bar{s}$, $bbbb\bar{n}$, and $bbbb\bar{s}$ subsystems in Fig. \ref{fig-QQQQq} (a)-(d), respectively.
Moreover, we also plot all the baryon-meson thresholds which they can decay to through quark rearrangement in Fig. \ref{fig-QQQQq} (a)-(d).
Meanwhile, we label the spin of the baryon-meson states with superscript.
When the spin of an initial pentaquark state is equal to the number in the superscript of a baryon-meson state,
it may decay into that baryon-meson channel through S wave.
Here, we define the relatively ``stable" pentaquark states as those which cannot decay into the S wave baryon-meson states.
Meanwhile, we label these stable pentaquark states with ``$\diamond$'' in the relevant figure and tables.

Based on the obtained $cccc\bar{q}$ and $bbbb\bar{q}$  pentaquark spectra,
we can discuss the possible decay patterns by considering different rearrangement of quarks in the corresponding pentaquark states. The discussion of possible decay patterns for these pentaquark states would be helpful for the observation in experiments.
From Fig. \ref{fig-QQQQq} (a)-(d), we can easily find that the $J^{P}=3/2^{-}$ state generally have smaller masses than that of the $J^{P}=1/2^{-}$ state in the $cccc\bar{q}$ and $bbbb\bar{q}$ pentaquark subsystems.
We also find that all the $cccc\bar{q}$ and $bbbb\bar{q}$ pentaquark states have strong decay channels, which indicates that in the modified CMI model, no stable $cccc\bar{q}$ and $bbbb\bar{q}$ pentaquark exists.

In addition to the mass spectra, the eigenvectors of pentaquark states will also provide important information about the two-body strong decay of multiquark states \cite{Jaffe:1976ig,Strottman:1979qu,Weng:2019ynva,Weng:2020jao,Zhao:2014qva,Wang:2015epa}.
The overlap for the pentaquark with a specific baryon $\otimes$ meson state can be calculated by transforming the eigenvectors of the pentaquark states into the baryon $\otimes$ meson configuration.
In the $QQQ\otimes Q\bar{q}$ configuration, the color wave function of the pentaquark falls into two categories:
the color-singlet $|(QQQ)^{1_{c}}(Q\bar{q})^{1_{c}}\rangle$ and the color-octet $|(QQQ)^{8_{c}}(Q\bar{q})^{8_{c}}\rangle$.
The former one can easily dissociate into an S-wave baryon and meson (the so-called Okubo-Zweig-Iizuka (OZI)-superallowed decays), while the latter one cannot fall apart without the gluon exchange force.
In this work, we only focus on the OZI-superallowed pentaquark decay process.
It means that only the $C_3$ color part in Eq.(\ref{eq-color4}) is considered in the color space.

For the two body decay via $L$-wave process, the expression describing partial decay width can be parameterized as \cite{Weng:2019ynva,Weng:2020jao}
\begin{eqnarray}\label{Eq20}
\Gamma_{i}=\gamma_{i}\alpha\frac{k^{2L+1}}{m^{2L}}\cdot|c_{i}|^{2},
\end{eqnarray}
where $\alpha$ is an effective coupling constant,
$m$ is the mass of the initial state,
$k$ is the momentum of the final states in the rest frame.
$c_i$ is the overlap between the pentaquark wave function and the meson + baryon wave function. Take $P_{c^{4}\bar{n}}(6761, 1/2, 3/2^{-})\to \Omega_{ccc} D^{*}$ as an example,
\begin{eqnarray}
c_i&\equiv& \langle\Omega_{ccc}\otimes D^{*}  |P_{c^{4}\bar{n}}(6761, \frac12, \frac32^{-})\rangle\\
&=&\langle(ccc)^{1_c}_{S=1/2}\otimes (c\bar n)^{1_c}_{S=1}  |P_{c^{4}\bar{n}}(6761, \frac12, \frac32^{-})\rangle=0.456. \nonumber\\
\end{eqnarray}
We show all possible overlaps between a pentaquark state and its possible $|(QQQ)^{1_{c}}(Q\bar{q})^{1_{c}}\rangle$ and $|(QQQ)^{1_{c}}(Q\bar{q})^{1_{c}}\rangle$ components in Table \ref{eigenvector-QQQQq}.
For the decay processes that we are interested in, $(k/m)^{2}$ is of $\mathcal{O}(10^{-2})$ or even smaller. Thus we only consider the $S$-wave decays.

As for the $\gamma_{i}$, it depends on the spatial wave functions of initial and final states, and may not be the same for different decay processes.
In the quark model, the spatial wave functions of the ground state scalar and vector meson are the same \cite{Weng:2019ynva}.
As a rough estimation, we introduce the following approximations to calculate the relative partial decay widths of the  $cccc\bar{q}$ and $bbbb\bar{q}$ pentaquark states:
\begin{eqnarray}\label{eq:gamma2}
\gamma_{\Omega_{ccc}D^{*}}&=\gamma_{\Omega_{ccc}D},\quad\gamma_{\Omega_{ccc}D_{s}^{*}}=&\gamma_{\Omega_{ccc}D_{s}},\nonumber\\
\gamma_{\Omega_{bbb}B^{*}}&=\gamma_{\Omega_{bbb}B},\quad\gamma_{\Omega_{bbb}B^{*}_{s}}=&\gamma_{\Omega_{bbb}B_{s}}.
\end{eqnarray}
We present $k\cdot|c_{i}|^{2}$ for each decay process in Table \ref{value-QQQQq}. From Table \ref{value-QQQQq}, one can roughly estimate the relative decay widths between different decay processes of different initial pentaquark states if neglecting the $\gamma_i$ differences.
Such approximation have already been applied in Refs. \cite{Cheng:2019obk,Cheng:2020nho}.
We emphasised that we are only interested in its relative partial decay widths between different decay modes for a particular pentaquark state.

Based on Eqs. (\ref{Eq20})-(\ref{eq:gamma2}), we can calculate relative partial decay widths for the $cccc\bar{q}$ and $bbbb\bar{q}$ pentaquark subsystems.
For the $J^{P}=1/2^{-}$ $cccc\bar{n}$ pentaquark state, it cannot decay into S-wave $\Omega_{ccc} D$ because of the constraint of angular conservation law.
Same situation also happen in the $cccc\bar{s}$, $bbbb\bar{n}$, and $bbbb\bar{s}$ subsystems.

Due to small phase spaces, the $\rm P_{c^4\bar{n}}(6761,1/2,3/2^-)$ and $\rm P_{c^4\bar{s}}(6864,0,3/2^-)$ can only decay into $\Omega_{ccc}D$ and $\Omega_{ccc}D_s$ final states, respectively. 
While for the two $J^{P}=3/2^{-}$ $bbbb\bar{q}$ pentaquark states, the $\rm P_{b^{4}\bar{n}}(19647,1/2,3/2^{-})$ and $\rm P_{b^{4}\bar{s}}(19736,0,3/2^{-})$, we find
\begin{equation}
\frac{\Gamma[P_{(b^{4}\bar{n})}(19647, 1/2, 3/2^{-})\rightarrow \Omega_{bbb}B]}{\Gamma[P_{(b^{4}\bar{n})}(19647, 1/2, 3/2^{-})\rightarrow \Omega_{bbb}B^{*}]}=1.3,
\end{equation}
and
\begin{equation}
\frac{\Gamma[P_{(b^{4}\bar{s})}(19736, 0, 3/2^{-})\rightarrow \Omega_{bbb}B_{s}]}{\Gamma[P_{(b^{4}\bar{s})}(19736, 0, 3/2^{-})\rightarrow \Omega_{bbb}B^{*}_{s}]}=1.4.
\end{equation}
respectively.
Thus the widths of the two modes do no differ very much.
\subsubsection{The $cccb\bar{q}$ and $bbbc\bar{q}$ pentaquark states}
Next we consider the $cccb\bar{q}$ and $bbbc\bar{q}$ pentaquark subsystems. The $cccb\bar{q}$ and $bbbc\bar{q}$ pentaquark subsystems include three identical heavy quarks.

The masses of $cccb\bar{q}$ and $bbbc\bar{q}$ pentaquark states can be determined in two schemes and shown in Table \ref{mass-QQQQq}.
In the reference mass scheme, we can exhaust two types of baryon-meson reference systems.
Specifically, we can use the $\Omega_{ccc}$+$B$ ($B_s$) and $\Omega_{ccb}$+$D$ ($D_s$) as reference systems to estimate the masses of the $cccb\bar{n}$ $(cccb\bar{s})$ pentaquark subsystem.
Similarly, the meson-baryon reference systems $\Omega_{bbb}$+$D$ ($D_s$) and $\Omega_{bbc}$+$B$ ($B_s$) are used to calculate the masses of the $bbbc\bar{n}$ $(bbbc\bar{s})$ pentaquark subsystem.
The obtained eigenvalues and masses of $cccb\bar{n}$ $(\bar{s})$ and $bbbc\bar{n}$ $(\bar{s})$ pentaquark states calculated from two types of reference systems are presented in the third and fourth columns of Table \ref{mass-QQQQq}, respectively.
We easily find that the mass spectra came from two different reference systems differ by more than 100 MeV for some studied subsystems.
The reason is that the dynamics and contributions from other terms in conventional meson and baryon potential are not elaborately considered in this model \cite{Zhou:2018pcv}.
However, the mass gaps under different reference systems are still same.
Thus, if one pentaquark state were observed, its partner states may be searched for with the relative positions presented in Table \ref{mass-QQQQq}.

Based on the results listed in the last column of Table \ref{mass-QQQQq}, we plot the mass spectra and relevant quark rearrangement decay patterns for the $cccb\bar{n}$, $cccb\bar{s}$, $bbbc\bar{n}$, and $bbbc\bar{s}$ subsystems in Fig. \ref{fig-QQQQq} (e)-(h), respectively.
Moreover, according to the modified CMI model, we can obtain the overlaps for $cccb\bar{n}$ $(cccb\bar{s})$ and $bbbc\bar{n}$ $(bbbc\bar{s})$ pentaquark states with different baryon $\otimes$ meson bases, and the results are shown in Table \ref{eigenvector-QQQQq}.

From Table \ref{eigenvector-QQQQq},
the $\rm P_{c^{3}b\bar{n}}(10110,1/2,5/2^{-})$ state couples completely to the $\Omega_{ccc}\bar{B}^{*}$ system , which can
be written as a direct product of a baryon $\Omega_{ccc}$ and a meson $\bar{B}^{*}$.
Moreover, for the $\rm P_{c^{3}b\bar{n}}(10118, 1/2, 3/2^{-})$, $\rm P_{c^{3}b\bar{n}}(10078, 1/2, 3/2^{-})$, and $\rm P_{c^{3}b\bar{n}}(10134, 1/2, 1/2^{-})$ states, they strongly couple to the $\Omega_{ccc}\bar{B}^{*}$, $\Omega_{ccc}\bar{B}$, and $\Omega_{ccc}\bar{B}^{*}$ bases, respectively.
This kind of pentaquark behaves similar to the ordinary scattering state made of a baryon and meson if the inner interaction is not strong, but could also be a resonance or bound state dynamically generated by the baryon and meson with strong interaction.
These kinds of pentaquarks deserve a more careful study with some hadron-hadron interaction models in future.
Thus, we label them with $``\star"$ in Tables \ref{eigenvector-QQQQq}, \ref{value-QQQQq}, and Fig. \ref{fig-QQQQq}.
Moreover, we find that the $J^{P}=5/2^{-}$ $QQQQ^{\prime}\bar{q}$ pentaquark states all have only one component $\Omega_{QQQ}B^{*}_{(s)}$ $(D^{*}_{(s)})$.
Therefore, the $J^{P}=5/2^{-}$ ground states are regarded as the states made of two hadrons.

For $cccb\bar{q}$ pentaquark states, they have two types of decay modes: $ccc-b\bar{q}$ and $ccb-c\bar{q}$.
Similarly, the $bbbc\bar{q}$ pentaquark states also have two types of decay modes: $bbb-c\bar{q}$ and $bbc-b\bar{q}$.
In the heavy quark limit, $\Omega^{*}_{ccb}$ ($\Omega^{*}_{bbc}$) and $\Omega_{ccb}$ ($\Omega_{bbc}$) have the same spatial wave function.
Thus, for a $cccb\bar{q}$ or $bbbc\bar{q}$ pentaquark state,
we use the following approximations
\begin{eqnarray}\label{eq:gamma3}
\gamma_{\Omega^{*}_{ccb}D^{*}}&=\gamma_{\Omega^{*}_{ccb}D}=\gamma_{\Omega_{ccb}D^{*}}=&\gamma_{\Omega_{ccb}D},\gamma_{\Omega_{ccc}B^{*}}=\gamma_{\Omega_{ccc}B},
\nonumber\\
\gamma_{\Omega^{*}_{ccb}D^{*}_{s}}&=\gamma_{\Omega^{*}_{ccb}D_{s}}=\gamma_{\Omega_{ccb}D^{*}_{s}}=&\gamma_{\Omega_{ccb}D_{s}},\gamma_{\Omega_{ccc}B^{*}_{s}}=\gamma_{\Omega_{ccc}B_{s}},
\nonumber\\
\gamma_{\Omega^{*}_{bbc}B^{*}}&=\gamma_{\Omega^{*}_{bbc}B}=\gamma_{\Omega_{bbc}B^{*}}=&\gamma_{\Omega_{bbc}B},
\gamma_{\Omega_{bbb}D^{*}}=\gamma_{\Omega_{bbb}D},\nonumber\\
\gamma_{\Omega^{*}_{bbc}B^{*}_{s}}&=\gamma_{\Omega^{*}_{bbc}B_{s}}=\gamma_{\Omega_{bbc}B^{*}_{s}}=&\gamma_{\Omega_{bbc}B_{s}}, \gamma_{\Omega_{bbb}D^{*}_{s}}=\gamma_{\Omega_{bbb}D_{s}}.\nonumber\\
\end{eqnarray}
Based on Table \ref{eigenvector-QQQQq}, we obtain $k\cdot|c_{i}|^{2}$ for each $cccb\bar{q}$  and $bbbc\bar{q}$  pentaquark state and present them in Table \ref{value-QQQQq}.

As an example, we only concentrate on the $cccb\bar{n}$ pentaquark subsystem. According to Table \ref{value-QQQQq}, we can see that the $\rm P_{c^{3}b\bar{n}}(9961, 1/2, 3/2^{-})$ and $\rm P_{c^{3}b\bar{n}}(9946, 1/2, 1/2^{-})$ states only decay into $\Omega^{*}_{ccb}D$ and $\Omega_{ccb}D$ final states, respectively.
For the $\rm P_{c^{3}b\bar{n}}(10062, 1/2, 1/2^{-})$ state, we have its relative partial decay width ratios as
\begin{equation}
\Gamma_{\Omega^{*}_{ccb}D^{*}}:\Gamma_{\Omega_{ccb}D^{*}}:\Gamma_{\Omega_{ccb}D}=24:24:1,
\end{equation}
which suggests that the partial decay width of the $\Omega^{*}_{ccb}D^{*}$ channel is nearly equal to that of the $\Omega_{ccb}D^{*}$ channel.
Note that if a state would be observed in the decay pattern $\Omega^{*}_{ccb}D^{*}$, $\Omega^{*}_{ccb}D$, $\Omega^{*}_{ccb}D^{*}$, and $\Omega_{ccb}D$ , it is a good candidate of a $cccb\bar{n}$ pentaquark state.

\subsubsection{The $ccbb\bar{q}$ pentaquark states}

The last group of the $QQQQ\bar{q}$ system is the pentaquark states with the $ccbb\bar{q}$ configuration.
The $ccbb\bar{q}$ pentaquark states have two pairs of identical heavy quarks, the $cc$ pair and $bb$ pair.
When we construct the wave functions of $ccbb\bar{q}$ pentaquark states, the Pauli principle should be satisfied simultaneously for these two pairs of heavy quarks.

In the reference mass scheme, there are also two types of meson-baryon reference systems for the $ccbb\bar{n}$ ($ccbb\bar{s}$) pentaquark subsystem, i.e., the $\Omega_{ccb}$+$B$ and $\Omega_{cbb}$+$D$ ($\Omega_{ccb}$+$B_s$ and $\Omega_{cbb}$+$D_s$).

Based on the results obtained from the modified CMI model in Table \ref{mass-QQQQq}, we plot the mass spectra and possible decay patterns via rearrangement of constituent quarks in $ccbb\bar{n}$ and $ccbb\bar{s}$ pentaquark states in Fig. \ref{fig-QQQQq} (i)-(j).
According to Fig. \ref{fig-QQQQq} (i)-(j),
we find that all $ccbb\bar{n}$ and $ccbb\bar{s}$ pentaquark states have strong decay channels.
i.e., from the modified CMI model analysis, there is no stable pentaquark state in $ccbb\bar{n}$ and $ccbb\bar{s}$ pentaquark subsystems.

To calculate the strong decay widths of the $ccbb\bar{n}$ and $ccbb\bar{s}$ pentaquark subsystems, we can use the following approximations
\begin{eqnarray}\label{eq:gamma4}
\gamma_{\Omega^{*}_{cbb}D^{*}}&=\gamma_{\Omega^{*}_{cbb}D}=\gamma_{\Omega_{cbb}D^{*}}=&\gamma_{\Omega_{cbb}D},\nonumber\\
\gamma_{\Omega^{*}_{ccb}B^{*}}&=\gamma_{\Omega^{*}_{ccb}B}=\gamma_{\Omega_{ccb}B^{*}}=&\gamma_{\Omega_{ccb}B},\nonumber\\
\gamma_{\Omega^{*}_{cbb}D_{s}^{*}}&=\gamma_{\Omega^{*}_{cbb}D_{s}}=\gamma_{\Omega_{cbb}D_{s}^{*}}=&\gamma_{\Omega_{cbb}D_{s}},\nonumber\\
\gamma_{\Omega^{*}_{ccb}B_{s}^{*}}&=\gamma_{\Omega^{*}_{ccb}B_{s}}=\gamma_{\Omega_{ccb}B_{s}^{*}}=&\gamma_{\Omega_{ccb}B_{s}}.
\end{eqnarray}
By introducing the above relations, $k\cdot|c_{i}|^{2}$ for $ccbb\bar{n}$ ($ccbb\bar{s}$) pentaquark states can be obtained and we present them in Table \ref{value-QQQQq}.

To discuss the strong decay behaviors of the $ccbb\bar{n}$ $(\bar{s})$ pentaquark states, we mainly focus on the relative partial decay widths of the $ccbb\bar{n}$ subsystem, and the $ccbb\bar{s}$ subsystem can be analyzed in a similar way.

From Table \ref{value-QQQQq}, the $J^{P}=5/2^{-}$, the lowest $J^{P}=3/2^{-}$, and the lowest $J^{P}=1/2^{-}$ states can only decay into $\Omega^{*}_{cbb}D^{*}$, $\Omega^{*}_{cbb}D$, and $\Omega_{cbb}D$, respectively.
The most important decay channel for the $\rm P_{c^{2}b^{2}\bar{n}}(13383, 1/2, 3/2^{-})$ is $\Omega^{*}_{bbc}D$ channel in $bbc-c\bar{n}$ decay mode.
The highest $J^P=3/2^-$ state $\rm P_{c^{2}b^{2}\bar{n}}(13414, 1/2, 1/2^{-})$ has many different decay channels and this state is expected to be broad.
For the $\rm P_{c^{2}b^{2}\bar{n}}(13383, 1/2, 3/2^{-})$, we have
\begin{equation}
\gamma_{\Omega^{*}_{cbb}D^{*}}:\gamma_{\Omega^{*}_{cbb}D}:\gamma_{\Omega_{cbb}D^{*}}=1.5:15:1,
\end{equation}
and
\begin{equation}
\gamma_{\Omega^{*}_{ccb}B^{*}}:\gamma_{\Omega^{*}_{ccb}B}:\gamma_{\Omega_{ccb}B^{*}}=1.5:1:1.4.
\end{equation}
For the $\rm P_{c^{2}b^{2}\bar{n}}(13414, 1/2, 1/2^{-})$, we have
\begin{equation}
\gamma_{\Omega^{*}_{cbb}D^{*}}:\gamma_{\Omega_{cbb}D^{*}}:\gamma_{\Omega_{cbb}D}=12:3.7:1,
\end{equation}
and
\begin{equation}
\gamma_{\Omega^{*}_{ccb}B^{*}}:\gamma_{\Omega_{ccb}B^{*}}:\gamma_{\Omega_{ccb}B}=3.5:0.4:1.
\end{equation}
Obviously, its dominant decay modes in $cbb-c\bar{n}$ and $ccb-b\bar{n}$ sectors are $\Omega^{*}_{cbb}D^{*}$ and $\Omega^{*}_{ccb}B^{*}$ channels, respectively.
Other three $J^{P}=3/2^{-}$ and three $J^{P}=1/2^{-}$  states only have $cbb-c\bar{n}$ decay mode.
The $ccb-b\bar{n}$ decay mode are strongly suppressed by the corresponding phase space.

\subsubsection{Comparison of other pentaquark systems}
In 2020, the LHCb collaboration studied the invariant mass spectrum of $J/\psi$ pairs, and they reported a narrow structure around 6.9 GeV \cite{Aaij:2020fnh}. Take this as an opportunity, the heavy flavored pentaquarks with four heavy quarks ($QQQQ\bar{q}$) and fully heavy pentaquarks ($QQQQ\bar{Q}$) are systematically discussed in this work and Ref. \cite{An:2020jix} within the modify CMI model. Here we can compare the differences between these two pentaquark systems.

Firstly, we discuss the mass differences among the ground states of $QQQQ\bar{n}$, $QQQQ\bar{s}$, $QQQQ\bar{c}$, and $QQQQ\bar{b}$ with the same $J^P$. From Table  \ref{eigenvector-QQQQq} and Table IV of
Ref. \cite{An:2020jix}, the masses of $cccc\bar{n}$, $cccc\bar{s}$, $cccc\bar{c}$, and $cccc\bar{b}$ ground states with $J^{P}=3/2^{-}$ are  6761, 6864, 7864, and 11130 MeV, respectively.
Moreover, relative to the the $J^{P}=1/2^{-}$ states, their corresponding mass gaps are 106 MeV, 108 MeV, 85 MeV, and 47 MeV, respectively.
Other subsystems also have similar situations.
Thus, compared with the $QQQQ\bar{Q}$ system, the $QQQQ\bar{q}$ system has lighter masses and bigger mass gaps when a heavy antiquark is replaced by a light antiquark because $v_{ij} \propto 1/{m_{i}m_{j}}$.

Secondly, we study the relations between the pentaquark and their corresponding baryon-meson channels.  The fully heavy pentaquarks $QQQQ\bar{Q}$ are more likely below all possible strong-decays channels and thus more stable compared to the $QQQQ\bar{q}$ systems. We have found two relatively stable states $\rm P_{c^{2}b^{2}\bar{b}}(17416, 0, 3/2^{-})$ and $\rm P_{c^{2}b^{2}\bar{b}}(17477, 0, 5/2^{-})$, which are below all allowed rearrangement decay channels in $QQQQ\bar{Q}$ system. However, we do not find any stable state for the $QQQQ\bar{q}$ multiquark systems. When both $QQQQ\bar{Q}$ and $QQQQ\bar{q}$ ground states are above their corresponding baryon-meson channel, the $QQQQ\bar{Q}$ mass would be closer to the threshold. For example,
the $cccc\bar{n}$, $cccc\bar{s}$, $cccc\bar{c}$, and $cccc\bar{b}$ states are above the corresponding minimum threshold ($\Omega_{ccc}$+ pseudoscalar meson) 106 MeV, 110 MeV, 94.5 MeV, and 69.5 MeV, respectively.

Unlike fully pentaquark $QQQQ\bar{Q}$, all $QQQQ\bar{q}$ pentaquark states can never mix with a triquark baryon and thus are explicit exotic states. Accurate measurement in future experiment and the comparison may help us understand the $Q\bar{Q}$ annihilation effects in the hadron spectrum.

\section{Summary}\label{sec5}
The observation of the $P_c(4312)$, $P_c(4440)$, and $P_c(4457)$ states achieved by the LHCb Collaboration and the study of the possible stable $QQ\bar{q}\bar{q}$ tetraquark states give us strong confidence to study the mass spectra of the $QQQQ\bar{q}$ pentaquark system within the framework of CMI model.
Similar to the fully-heavy $QQ\bar{Q}\bar{Q}$ tetraquark system \cite{Aaij:2020fnh}, the $QQQQ\bar{q}$ system consist of four heavy quarks are dominantly bounded by the gluon exchange interaction, and can hardly be considered as molecular states.

In this work, by including the flavor SU(3) breaking effect, we firstly construct the $\psi_{flavor}\otimes\psi_{color}\otimes\psi_{spin}$ wave functions based on the SU(2) and SU(3) symmetry and Pauli Principle.
Then we extract the effective coupling constants from the conventional hadrons. After that, we systematically calculate the chromomagnetic interaction Hamiltonian for the discussed pentaquark states and obtain the corresponding mass spectra.
As a modification to the CMI model, the effect of chromoelectric interaction is added in the modified CMI model.
So, we mainly discussed the results of mass spectra for the $QQQQ\bar{q}$ pentaquark system obtained from the modified CMI model. The results from the reference mass scheme are presented for comparison.
In addition to the eigenvalues, we also provide the eigenvectors to extract useful information about the decay properties for the studied pentaquark systems.
Finally, we analyze the stability, possible quark rearrangement decay channels and relative partial decay widths for all the $QQQQ\bar{q}$ pentaquark states.

Due to the constraint from Pauli principle, there is no ground $J^{P}=5/2^{-}$ $cccc\bar{q}$ and $bbbb\bar{q}$ pentaquark states.
From the obtained Tables and Figs for the $QQQQ\bar{q}$ pentaquark system, we find no stable candidate
in the modified CMI model.
However, due to the uncertainty of the modified CMI model, some of them may not truly be unstable states, and further dynamical calculations may help us to clarify their natures.
Especially, for some unstable states which are a little higher than the meson-baryon thresholds of lowest strong decay channels, they can be considered as narrow pentaquark states, and have opportunities to be found in future experiment.
Meanwhile, the whole mass spectra has a slight shift or down due to the mass deviation of constituent quarks.
While the mass gaps between different pentaquark states are relatively stable, if one pentaquark states are observed in experiment, we can use these mass gaps to predict their corresponding multiplets.


Among the studied $QQQQ\bar{q}$ pentaquark states, all of them are explicit exotic states. If such pentaquark states
are observed, their exotic nature can be easily identified.
However, up to now, none of $QQQQ\bar{q}$ pentaquark states is found.
More detailed dynamical investigations on these pentaquark systems are still needed.
Producing a $QQQQ\bar{q}$ pentaquark state seems to be a difficult task in experiment.
Our systematical study may provide theorists and experimentalists some preliminary understanding toward this pentaquark system.
We hope that the present study may inspire experimentalists and theorists to pay attention to this kind of pentaquark system.

\section*{ACKNOWLEDGMENTS}
This work is supported by the China National Funds for Distinguished Young Scientists under Grant No. 11825503, National Key Research and Development Program of China under Contract No. 2020YFA0406400, the 111 Project under Grant No. B20063, and the National Natural Science Foundation of China under Grant No. 12047501. This project is also supported by the National Natural Science Foundation of China under Grants No. 11705072, 11705069, and 11965016, CAS Interdisciplinary Innovation Team.

\begin{widetext}
\appendix
\section{Some expressions in detail}\label{sec10}

The possible color $\otimes$ spin Young-Yamanouchi basis vectors of the Young tableaux in Eq. (\ref{colorspin}) are presented in Eq. (\ref{colorspin1}). We list the possible wave function satisfied by Pauli Principle in Table \ref{flavor} and some CMI Hamiltonians in Table. \ref{nnnsQ}.

\begin{align}
J=\frac{5}{2}: \quad
&\begin{tabular}{|c|c|}
\hline
1 &3     \\
\cline{1-2}
2      \\
\cline{1-1}
4      \\
\cline{1-1}
\end{tabular}_{CS_1}
=
\begin{tabular}{|c|c|}
\hline
1 &3    \\
\cline{1-2}
2     \\
\cline{1-1}
4      \\
\cline{1-1}
\end{tabular}_{C_2}
\otimes
\begin{tabular}{|c|c|c|c|}
\hline
1&2&3&4  \\
\cline{1-4}
\end{tabular}_{S_1};
\nonumber
\\
&\begin{tabular}{|c|c|}
\hline
1 &4     \\
\cline{1-2}
2      \\
\cline{1-1}
3      \\
\cline{1-1}
\end{tabular}_{CS_2}
=
\begin{tabular}{|c|c|}
\hline
1 &4    \\
\cline{1-2}
2     \\
\cline{1-1}
3      \\
\cline{1-1}
\end{tabular}_{C_3}
\otimes
\begin{tabular}{|c|c|c|c|}
\hline
1&2&3&4  \\
\cline{1-4}
\end{tabular}_{S_1};
\nonumber
\end{align}
\begin{align}\label{colorspin1}
J=\frac{3}{2}: \quad
&\begin{tabular}{|c|}
\hline
1      \\
\cline{1-1}
2      \\
\cline{1-1}
3      \\
\cline{1-1}
4      \\
\cline{1-1}
\end{tabular}_{CS_1}
=
\frac{1}{\sqrt{3}}
\begin{tabular}{|c|c|}
\hline
1 &2     \\
\cline{1-2}
3      \\
\cline{1-1}
4      \\
\cline{1-1}
\end{tabular}_{C_1}
\otimes
\begin{tabular}{|c|c|c|}
\hline
1 &3&4     \\
\cline{1-3}
2      \\
\cline{1-1}
\end{tabular}_{S_5}
-\frac{1}{\sqrt{3}}
\begin{tabular}{|c|c|}
\hline
1 &3     \\
\cline{1-2}
2      \\
\cline{1-1}
4      \\
\cline{1-1}
\end{tabular}_{C_2}
\otimes
\begin{tabular}{|c|c|c|}
\hline
1 &2&4     \\
\cline{1-3}
3      \\
\cline{1-1}
\end{tabular}_{S_4}
+\frac{1}{\sqrt{3}}
\begin{tabular}{|c|c|}
\hline
1 &4    \\
\cline{1-2}
2      \\
\cline{1-1}
3      \\
\cline{1-1}
\end{tabular}_{C_3}
\otimes
\begin{tabular}{|c|c|c|}
\hline
1 &2&3     \\
\cline{1-3}
4      \\
\cline{1-1}
\end{tabular}_{S_3};
\nonumber
\\
&\begin{tabular}{|c|c|}
\hline
1 &4     \\
\cline{1-2}
2      \\
\cline{1-1}
3      \\
\cline{1-1}
\end{tabular}_{CS_2}
=
-\frac{1}{\sqrt{6}}
\begin{tabular}{|c|c|}
\hline
1 &2     \\
\cline{1-2}
3      \\
\cline{1-1}
4      \\
\cline{1-1}
\end{tabular}_{C_1}
\otimes
\begin{tabular}{|c|c|c|}
\hline
1 &3&4     \\
\cline{1-3}
2      \\
\cline{1-1}
\end{tabular}_{S_5}
+\frac{1}{\sqrt{6}}
\begin{tabular}{|c|c|}
\hline
1 &3     \\
\cline{1-2}
2      \\
\cline{1-1}
4      \\
\cline{1-1}
\end{tabular}_{C_2}
\otimes
\begin{tabular}{|c|c|c|}
\hline
1 &2&4     \\
\cline{1-3}
3      \\
\cline{1-1}
\end{tabular}_{S_4}
+{\sqrt{\frac{2}{3}}}
\begin{tabular}{|c|c|}
\hline
1 &4    \\
\cline{1-2}
2      \\
\cline{1-1}
3      \\
\cline{1-1}
\end{tabular}_{C_3}
\otimes
\begin{tabular}{|c|c|c|}
\hline
1 &2&3     \\
\cline{1-3}
4      \\
\cline{1-1}
\end{tabular}_{S_3};
\nonumber
\\
&\begin{tabular}{|c|c|}
\hline
1 &3     \\
\cline{1-2}
2     \\
\cline{1-1}
4     \\
\cline{1-1}
\end{tabular}_{CS_3}
=
\frac{1}{\sqrt3}
\begin{tabular}{|c|c|}
\hline
1 &2     \\
\cline{1-2}
3     \\
\cline{1-1}
4      \\
\cline{1-1}
\end{tabular}_{C_1}
\otimes
\begin{tabular}{|c|c|c|}
\hline
1 &3&4    \\
\cline{1-3}
2\\
\cline{1-1}
\end{tabular}_{S_5}
-\frac{1}{\sqrt6}
\begin{tabular}{|c|c|}
\hline
1 &3     \\
\cline{1-2}
2      \\
\cline{1-1}
4      \\
\cline{1-1}
\end{tabular}_{C_2}
\otimes
\begin{tabular}{|c|c|c|}
\hline
1 &2&3    \\
\cline{1-3}
4\\
\cline{1-1}
\end{tabular}_{S_3}
+\frac{1}{\sqrt3}
\begin{tabular}{|c|c|}
\hline
1 &3     \\
\cline{1-2}
2     \\
\cline{1-1}
4      \\
\cline{1-1}
\end{tabular}_{C_2}
\otimes
\begin{tabular}{|c|c|c|}
\hline
1 &2&4    \\
\cline{1-3}
3\\
\cline{1-1}
\end{tabular}_{S_4}
+\frac{1}{\sqrt6}
\begin{tabular}{|c|c|}
\hline
1 &4     \\
\cline{1-2}
2     \\
\cline{1-1}
3      \\
\cline{1-1}
\end{tabular}_{C_3}
\otimes
\begin{tabular}{|c|c|c|}
\hline
1 &2&4    \\
\cline{1-3}
3\\
\cline{1-1}
\end{tabular}_{S_4};
\nonumber
\\
&\begin{tabular}{|c|c|}
\hline
1 & 3     \\
\cline{1-2}
2 & 4     \\
\cline{1-2}
\end{tabular}_{CS_4}
=
-\frac{1}{\sqrt6}
\begin{tabular}{|c|c|}
\hline
1 &2     \\
\cline{1-2}
3     \\
\cline{1-1}
4      \\
\cline{1-1}
\end{tabular}_{C_1}
\otimes
\begin{tabular}{|c|c|c|}
\hline
1 &3&4    \\
\cline{1-3}
2\\
\cline{1-1}
\end{tabular}_{S_5}
-\frac{1}{\sqrt3}
\begin{tabular}{|c|c|}
\hline
1 &3     \\
\cline{1-2}
2      \\
\cline{1-1}
4      \\
\cline{1-1}
\end{tabular}_{C_2}
\otimes
\begin{tabular}{|c|c|c|}
\hline
1 &2&3    \\
\cline{1-3}
4\\
\cline{1-1}
\end{tabular}_{S_3}
-\frac{1}{\sqrt6}
\begin{tabular}{|c|c|}
\hline
1 &3     \\
\cline{1-2}
2     \\
\cline{1-1}
4      \\
\cline{1-1}
\end{tabular}_{C_2}
\otimes
\begin{tabular}{|c|c|c|}
\hline
1 &2&4    \\
\cline{1-3}
3\\
\cline{1-1}
\end{tabular}_{S_4}
+\frac{1}{\sqrt3}
\begin{tabular}{|c|c|}
\hline
1 &4     \\
\cline{1-2}
2     \\
\cline{1-1}
3      \\
\cline{1-1}
\end{tabular}_{C_3}
\otimes
\begin{tabular}{|c|c|c|}
\hline
1 &2&4    \\
\cline{1-3}
3\\
\cline{1-1}
\end{tabular}_{S_4};
\nonumber
\\
&\begin{tabular}{|c|c|}
\hline
1 &4     \\
\cline{1-2}
2      \\
\cline{1-1}
3      \\
\cline{1-1}
\end{tabular}_{CS_{5}}
=
\begin{tabular}{|c|c|}
\hline
1 &4    \\
\cline{1-2}
2     \\
\cline{1-1}
3      \\
\cline{1-1}
\end{tabular}_{C_3}
\otimes
\begin{tabular}{|c|c|c|c|}
\hline
1&2&3&4  \\
\cline{1-4}
\end{tabular}_{S_2};
\nonumber
\\
&\begin{tabular}{|c|c|}
\hline
1 &3     \\
\cline{1-2}
2      \\
\cline{1-1}
4      \\
\cline{1-1}
\end{tabular}_{CS_{6}}
=
\begin{tabular}{|c|c|}
\hline
1 &3    \\
\cline{1-2}
2     \\
\cline{1-1}
4     \\
\cline{1-1}
\end{tabular}_{C_2}
\otimes
\begin{tabular}{|c|c|c|c|}
\hline
1&2&3&4  \\
\cline{1-4}
\end{tabular}_{S_2}.
\nonumber
\\
J=\frac12: \quad
&\begin{tabular}{|c|}
\hline
1      \\
\cline{1-1}
2      \\
\cline{1-1}
3      \\
\cline{1-1}
4      \\
\cline{1-1}
\end{tabular}_{CS_1}
=
\frac{1}{\sqrt3}
\begin{tabular}{|c|c|}
\hline
1 &2     \\
\cline{1-2}
3      \\
\cline{1-1}
4      \\
\cline{1-1}
\end{tabular}_{C_1}
\otimes
\begin{tabular}{|c|c|c|}
\hline
1 &3&4     \\
\cline{1-3}
2      \\
\cline{1-1}
\end{tabular}_{S_8}
-\frac{1}{\sqrt3}
\begin{tabular}{|c|c|}
\hline
1 &3     \\
\cline{1-2}
2      \\
\cline{1-1}
4      \\
\cline{1-1}
\end{tabular}_{C_2}
\otimes
\begin{tabular}{|c|c|c|}
\hline
1 &2&4     \\
\cline{1-3}
3      \\
\cline{1-1}
\end{tabular}_{S_7}
+\frac{1}{\sqrt3}
\begin{tabular}{|c|c|}
\hline
1 &4    \\
\cline{1-2}
2      \\
\cline{1-1}
3      \\
\cline{1-1}
\end{tabular}_{C_3}
\otimes
\begin{tabular}{|c|c|c|}
\hline
1 &2&3     \\
\cline{1-3}
4      \\
\cline{1-1}
\end{tabular}_{S_6};
\nonumber
\\
&\begin{tabular}{|c|c|}
\hline
1 &4     \\
\cline{1-2}
2      \\
\cline{1-1}
3      \\
\cline{1-1}
\end{tabular}_{CS_2}
=
-\frac{1}{\sqrt6}
\begin{tabular}{|c|c|}
\hline
1 &2     \\
\cline{1-2}
3      \\
\cline{1-1}
4      \\
\cline{1-1}
\end{tabular}_{C_1}
\otimes
\begin{tabular}{|c|c|c|}
\hline
1 &3&4     \\
\cline{1-3}
2      \\
\cline{1-1}
\end{tabular}_{S_8}
+\frac{1}{\sqrt6}
\begin{tabular}{|c|c|}
\hline
1 &3     \\
\cline{1-2}
2      \\
\cline{1-1}
4      \\
\cline{1-1}
\end{tabular}_{C_2}
\otimes
\begin{tabular}{|c|c|c|}
\hline
1&2 &4     \\
\cline{1-3}
3      \\
\cline{1-1}
\end{tabular}_{S_7}
+\sqrt{\frac23}
\begin{tabular}{|c|c|}
\hline
1 &4    \\
\cline{1-2}
2      \\
\cline{1-1}
3     \\
\cline{1-1}
\end{tabular}_{C_3}
\otimes
\begin{tabular}{|c|c|c|}
\hline
1 &2 &3    \\
\cline{1-3}
4      \\
\cline{1-1}
\end{tabular}_{S_6};
\nonumber
\\
&\begin{tabular}{|c|c|}
\hline
1 &3     \\
\cline{1-2}
2      \\
\cline{1-1}
4      \\
\cline{1-1}
\end{tabular}_{CS_3}
=
\frac{1}{\sqrt3}
\begin{tabular}{|c|c|}
\hline
1 &2     \\
\cline{1-2}
3      \\
\cline{1-1}
4      \\
\cline{1-1}
\end{tabular}_{C_1}
\otimes
\begin{tabular}{|c|c|c|}
\hline
1 &3&4     \\
\cline{1-3}
2      \\
\cline{1-1}
\end{tabular}_{S_8}
-\frac{1}{\sqrt6}
\begin{tabular}{|c|c|}
\hline
1 &3     \\
\cline{1-2}
2      \\
\cline{1-1}
4      \\
\cline{1-1}
\end{tabular}_{C_2}
\otimes
\begin{tabular}{|c|c|c|}
\hline
1&2 &3     \\
\cline{1-3}
4      \\
\cline{1-1}
\end{tabular}_{S_6}
+\frac{1}{\sqrt3}
\begin{tabular}{|c|c|}
\hline
1 &3     \\
\cline{1-2}
2      \\
\cline{1-1}
4      \\
\cline{1-1}
\end{tabular}_{C_2}
\otimes
\begin{tabular}{|c|c|c|}
\hline
1 &3 &4    \\
\cline{1-3}
2      \\
\cline{1-1}
\end{tabular}_{S_7}
+\frac{1}{\sqrt6}
\begin{tabular}{|c|c|}
\hline
1 &4     \\
\cline{1-2}
2      \\
\cline{1-1}
3      \\
\cline{1-1}
\end{tabular}_{C_3}
\otimes
\begin{tabular}{|c|c|c|}
\hline
1 &3 &4    \\
\cline{1-3}
2      \\
\cline{1-1}
\end{tabular}_{S_7};
\nonumber
\\
&\begin{tabular}{|c|c|}
\hline
1 &3     \\
\cline{1-2}
2 &4     \\
\cline{1-2}
\end{tabular}_{CS_4}
=
-\frac{1}{\sqrt6}
\begin{tabular}{|c|c|}
\hline
1 &2     \\
\cline{1-2}
3     \\
\cline{1-1}
4      \\
\cline{1-1}
\end{tabular}_{C_1}
\otimes
\begin{tabular}{|c|c|c|}
\hline
1 &3&4    \\
\cline{1-3}
2\\
\cline{1-1}
\end{tabular}_{S_8}
-\frac{1}{\sqrt3}
\begin{tabular}{|c|c|}
\hline
1 &3     \\
\cline{1-2}
2      \\
\cline{1-1}
4      \\
\cline{1-1}
\end{tabular}_{C_2}
\otimes
\begin{tabular}{|c|c|c|}
\hline
1 &2&3    \\
\cline{1-3}
4\\
\cline{1-1}
\end{tabular}_{S_6}
-\frac{1}{\sqrt6}
\begin{tabular}{|c|c|}
\hline
1 &3     \\
\cline{1-2}
2     \\
\cline{1-1}
4      \\
\cline{1-1}
\end{tabular}_{C_2}
\otimes
\begin{tabular}{|c|c|c|}
\hline
1 &2&4    \\
\cline{1-3}
3\\
\cline{1-1}
\end{tabular}_{S_7}
+\frac{1}{\sqrt3}
\begin{tabular}{|c|c|}
\hline
1 &4     \\
\cline{1-2}
2     \\
\cline{1-1}
3      \\
\cline{1-1}
\end{tabular}_{C_3}
\otimes
\begin{tabular}{|c|c|c|}
\hline
1 &2&4    \\
\cline{1-3}
3\\
\cline{1-1}
\end{tabular}_{S_7};
\nonumber
\\
&\begin{tabular}{|c|c|}
\hline
1 &4     \\
\cline{1-2}
2      \\
\cline{1-1}
3      \\
\cline{1-1}
\end{tabular}_{CS_{5}}
=
\frac{1}{\sqrt2}
\begin{tabular}{|c|c|}
\hline
1 &2     \\
\cline{1-2}
3      \\
\cline{1-1}
4      \\
\cline{1-1}
\end{tabular}_{C_1}
\otimes
\begin{tabular}{|c|c|}
\hline
1&3     \\
\cline{1-2}
2&4     \\
\cline{1-2}
\end{tabular}_{S_{10}}
-\frac{1}{\sqrt2}
\begin{tabular}{|c|c|}
\hline
1 &3     \\
\cline{1-2}
2      \\
\cline{1-1}
4      \\
\cline{1-1}
\end{tabular}_{C_2}
\otimes
\begin{tabular}{|c|c|}
\hline
1&2     \\
\cline{1-2}
3&4     \\
\cline{1-2}
\end{tabular}_{S_9};
\nonumber
\\
&\begin{tabular}{|c|c|}
\hline
1 &3     \\
\cline{1-2}
2      \\
\cline{1-1}
4      \\
\cline{1-1}
\end{tabular}_{CS_{6}}
=
\frac12
\begin{tabular}{|c|c|}
\hline
1 &2     \\
\cline{1-2}
3      \\
\cline{1-1}
4      \\
\cline{1-1}
\end{tabular}_{C_1}
\otimes
\begin{tabular}{|c|c|}
\hline
1&3     \\
\cline{1-2}
2&4     \\
\cline{1-2}
\end{tabular}_{S_{10}}
+\frac12
\begin{tabular}{|c|c|}
\hline
1 &3     \\
\cline{1-2}
2      \\
\cline{1-1}
4      \\
\cline{1-1}
\end{tabular}_{C_2}
\otimes
\begin{tabular}{|c|c|}
\hline
1&2     \\
\cline{1-2}
3&4     \\
\cline{1-2}
\end{tabular}_{S_9}
-\frac{1}{\sqrt2}
\begin{tabular}{|c|c|}
\hline
1 &4     \\
\cline{1-2}
2      \\
\cline{1-1}
3      \\
\cline{1-1}
\end{tabular}_{C_3}
\otimes
\begin{tabular}{|c|c|}
\hline
1&2     \\
\cline{1-2}
3&4     \\
\cline{1-2}
\end{tabular}_{S_9};
\nonumber
\\
\end{align}
\end{widetext}

\begin{table*}[t]
\caption{All possible color $\otimes$ spin Young-Yamanouchi basis vectors satisfied with Pauli Principle for a specific $QQQQ\bar{q}$ [$q=n,\,s$ ($n=u$, $d$) and $Q=c,\,b$] subsystem. The $J$ represents the spin of the pentaquark states.}
\label{flavor} \centering
\renewcommand\arraystretch{0.94}
\begin{tabular}{|c|c|c|}
\midrule[1.5pt]
\bottomrule[0.5pt]
Flavor waves&$J$&{The color $\otimes$ spin Young-Yamanouchi basis vectors}\\
\hline
\multirow{8}*{\makecell[c]{$cccc\bar{q}$\\  \\$bbbb\bar{q}$}}&$J=3/2$&
\begin{tabular}{|c|c|c}
\cline{1-1}
1 \\
\cline{1-1}
2\\
\cline{1-1}
3\\
\cline{1-1}
4 &\multicolumn{1}{c}{\quad}&$CS_1$\\
\cline{1-1}
\end{tabular}
\\
&&\\
&$J=1/2$&
\begin{tabular}{|c|c|c}
\cline{1-1}
1 \\
\cline{1-1}
2\\
\cline{1-1}
3\\
\cline{1-1}
4 &\multicolumn{1}{c}{\quad}&$CS_1$\\
\cline{1-1}
\end{tabular}
\\
&&\\
\cline{1-3}
\multirow{10}*{\makecell[c]{$cccb\bar{q}$\\  \\$bbbc\bar{q}$}}&$J=5/2$&
\begin{tabular}{|c|c|c}
\cline{1-2}
1 &4\\
\cline{1-2}
2\\
\cline{1-1}
3 &\multicolumn{1}{c}{\quad}&$CS_1$\\
\cline{1-1}
\end{tabular}
\\
&&\\
&$J=3/2$&
\begin{tabular}{|c|c|c}
\cline{1-1}
1 \\
\cline{1-1}
2\\
\cline{1-1}
3\\
\cline{1-1}
4 &\multicolumn{1}{c}{\quad}&$CS_1$\\
\cline{1-1}
\end{tabular}
,
\begin{tabular}{|c|c|c}
\cline{1-2}
1 &4\\
\cline{1-2}
2\\
\cline{1-1}
3 &\multicolumn{1}{c}{\quad}&$CS_2$\\
\cline{1-1}
\end{tabular}
,
\begin{tabular}{|c|c|c}
\cline{1-2}
1 &4\\
\cline{1-2}
2\\
\cline{1-1}
3 &\multicolumn{1}{c}{\quad}&$CS_5$\\
\cline{1-1}
\end{tabular}
\\
&&\\

&$J=1/2$&
\begin{tabular}{|c|c|c}
\cline{1-1}
1 \\
\cline{1-1}
2\\
\cline{1-1}
3\\
\cline{1-1}
4 &\multicolumn{1}{c}{\quad}&$CS_1$\\
\cline{1-1}
\end{tabular}
,
\begin{tabular}{|c|c|c}
\cline{1-2}
1 &4\\
\cline{1-2}
2\\
\cline{1-1}
3 &\multicolumn{1}{c}{\quad}&$CS_2$\\
\cline{1-1}
\end{tabular}
,
\begin{tabular}{|c|c|c}
\cline{1-2}
1 &4\\
\cline{1-2}
2\\
\cline{1-1}
3 &\multicolumn{1}{c}{\quad}&$CS_5$\\
\cline{1-1}
\end{tabular}
\\
&&\\
\cline{1-3}
\multirow{10}*{$ccbb\bar{q}$}&$J=5/2$&
\begin{tabular}{|c|c|c}
\cline{1-2}
1 &4\\
\cline{1-2}
2\\
\cline{1-1}
3 &\multicolumn{1}{c}{\quad}&$CS_1$\\
\cline{1-1}
\end{tabular}
\\
&&\\
&$J=3/2$&
\begin{tabular}{|c|c|c}
\cline{1-1}
1 \\
\cline{1-1}
2\\
\cline{1-1}
3\\
\cline{1-1}
4 &\multicolumn{1}{c}{\quad}&$CS_1$\\
\cline{1-1}
\end{tabular}
,
$\sqrt{\frac23}$
\begin{tabular}{|c|c|c}
\cline{1-2}
1 &3\\
\cline{1-2}
2\\
\cline{1-1}
4 &\multicolumn{1}{c}{\quad}&$CS_3$\\
\cline{1-1}
\end{tabular}
$-\sqrt{\frac13}$
\begin{tabular}{|c|c|c}
\cline{1-2}
1 &4\\
\cline{1-2}
2\\
\cline{1-1}
3 &\multicolumn{1}{c}{\quad}&$CS_2$\\
\cline{1-1}
\end{tabular}
,
\begin{tabular}{|c|c|c}
\cline{1-2}
1 &3&\\
\cline{1-2}
2 &4&$CS_4$\\
\cline{1-2}
\end{tabular}
,
$\sqrt{\frac23}$
\begin{tabular}{|c|c|c}
\cline{1-2}
1 &3\\
\cline{1-2}
2\\
\cline{1-1}
4 &\multicolumn{1}{c}{\quad}&$CS_6$\\
\cline{1-1}
\end{tabular}
$-\sqrt{\frac13}$
\begin{tabular}{|c|c|c}
\cline{1-2}
1 &4\\
\cline{1-2}
2\\
\cline{1-1}
3 &\multicolumn{1}{c}{\quad}&$CS_5$\\
\cline{1-1}
\end{tabular}
\\
&&\\

&$J=1/2$&
\begin{tabular}{|c|c|c}
\cline{1-1}
1 \\
\cline{1-1}
2\\
\cline{1-1}
3\\
\cline{1-1}
4 &\multicolumn{1}{c}{\quad}&$CS_1$\\
\cline{1-1}
\end{tabular}
,
$\sqrt{\frac23}$
\begin{tabular}{|c|c|c}
\cline{1-2}
1 &3\\
\cline{1-2}
2\\
\cline{1-1}
4 &\multicolumn{1}{c}{\quad}&$CS_3$\\
\cline{1-1}
\end{tabular}
$-\sqrt{\frac13}$
\begin{tabular}{|c|c|c}
\cline{1-2}
1 &4\\
\cline{1-2}
2\\
\cline{1-1}
3 &\multicolumn{1}{c}{\quad}&$CS_2$\\
\cline{1-1}
\end{tabular}
,
\begin{tabular}{|c|c|c}
\cline{1-2}
1 &3&\\
\cline{1-2}
2 &4&$CS_4$\\
\cline{1-2}
\end{tabular}
,
$\sqrt{\frac23}$
\begin{tabular}{|c|c|c}
\cline{1-2}
1 &3\\
\cline{1-2}
2\\
\cline{1-1}
4 &\multicolumn{1}{c}{\quad}&$CS_6$\\
\cline{1-1}
\end{tabular}
$-\sqrt{\frac13}$
\begin{tabular}{|c|c|c}
\cline{1-2}
1 &4\\
\cline{1-2}
2\\
\cline{1-1}
3 &\multicolumn{1}{c}{\quad}&$CS_5$\\
\cline{1-1}
\end{tabular}
\\
\bottomrule[0.5pt]
\midrule[1.5pt]
\end{tabular}
\end{table*}

\newsavebox{\tablebox}
\begin{table*}[t]
\caption{The expressions of CMI Hamiltonians for $cccc\bar{n}$, $cccb\bar{n}$, and $ccbb\bar{n}$ pentaquark subsystems. The $J$ represents the spin of the pentaquark states.} \label{nnnsQ} \centering
\begin{lrbox}{\tablebox}
\renewcommand\arraystretch{1.0}
\begin{tabular}{|c|c|}
\midrule[1.5pt]
\bottomrule[0.5pt]
$J$&The expressions of CMI Hamiltonian for $cccc\bar{n}$ subsystem \\
\hline
$J=3/2$&$\frac{56}{3}C_{cc}-\frac{16}{3}C_{c\bar{n}}$\\
$J=1/2$&$\frac{56}{3}C_{cc}+\frac{32}{3}C_{c\bar{n}}$\\ \cline{1-2}
$J$&The expressions of CMI Hamiltonian for $cccb\bar{n}$ subsystem \\ \cline{1-2}
$J=5/2$&$8C_{cc}+\frac{16}{3}C_{b\bar{n}}$\\
$J=3/2$&$\begin{pmatrix}
\begin{pmatrix}\frac{28}{3}C_{cc}+\frac{28}{3}C_{cb}\\-4C_{c\bar{n}}-\frac{4}{3}C_{b\bar{n}}\end{pmatrix}&
\frac{2\sqrt{2}}{3}\begin{pmatrix}-C_{cc}+C_{cb}\\+C_{c\bar{b}}-C_{b\bar{n}}\end{pmatrix}&
-\frac{8\sqrt{5}}{3}(C_{c\bar{n}}-C_{b\bar{n}})\\

\frac{2\sqrt{2}}{3}\begin{pmatrix}-C_{cc}+C_{cb}\\+C_{c\bar{b}}-C_{b\bar{n}}\end{pmatrix}&
\begin{pmatrix}\frac{26}{3}C_{cc}-6C_{cb}\\+\frac23C_{c\bar{b}}-2C_{b\bar{n}}\end{pmatrix}&
\frac{4\sqrt{10}}{3}(C_{c\bar{n}}+2C_{b\bar{n}})\\

-\frac{8\sqrt{5}}{3}(C_{c\bar{n}}-C_{b\bar{n}})&
\frac{4\sqrt{10}}{3}(C_{c\bar{n}}+2C_{b\bar{n}})&
8(C_{cc}-C_{b\bar{n}})
\end{pmatrix}$\\
$J=1/2$&$\begin{pmatrix}
\begin{pmatrix}\frac{28}{3}C_{cc}+\frac{28}{3}C_{cb}\\+8C_{c\bar{n}}+\frac83C_{b\bar{n}}\end{pmatrix}&
\frac{2\sqrt{2}}{3}\begin{pmatrix}-C_{cc}+C_{cb}\\-2C_{c\bar{n}}+2C_{b\bar{n}}\end{pmatrix}&
\frac{2\sqrt{2}}{3}(C_{c\bar{n}}-C_{b\bar{n}})\\

\frac{2\sqrt{2}}{3}\begin{pmatrix}-C_{cc}+C_{cb}\\-2C_{c\bar{n}}+2C_{b\bar{n}}\end{pmatrix}&
\begin{pmatrix}\frac{26}{3}C_{cc}-6C_{cb}\\-\frac43C_{c\bar{n}}+4C_{b\bar{n}}\end{pmatrix}&
-\frac{2}{3}(13C_{c\bar{n}}-C_{b\bar{n}})\\

\frac{2\sqrt{2}}{3}(C_{c\bar{n}}-C_{b\bar{n}})&
-\frac{2}{3}(13C_{c\bar{n}}-C_{b\bar{n}})&
10(C_{cc}-C_{cb})
\end{pmatrix}$\\
\cline{1-2}
\Xcline{1-2}{0.7pt}
$J$&The expressions of CMI Hamiltonian for $ccbb\bar{n}$ subsystem \\ \cline{1-2}
$J=5/2$&$\frac83(C_{cc}+C_{bb}+C_{cb}+C_{c\bar{b}}+C_{b\bar{n}})$\\
$J=3/2$&$\begin{pmatrix}
\begin{pmatrix}\frac{28}{9}C_{cc}+\frac{28}{9}C_{bb}+\frac{112}{9}C_{cb}\\-\frac83C_{c\bar{n}}-\frac83C_{b\bar{n}}\end{pmatrix}&
\frac{2}{3}\sqrt{\frac23}\begin{pmatrix}C_{cc}-C_{bb}\\-2C_{c\bar{n}}+2C_{b\bar{n}}\end{pmatrix}&
-\frac{2\sqrt{2}}{9}\begin{pmatrix}C_{cc}+C_{bb}+2C_{cb}\end{pmatrix}&
\frac{16}{3}\sqrt{\frac53}(C_{c\bar{n}}-C_{b\bar{n}})\\

\frac{2}{3}\sqrt{\frac23}\begin{pmatrix}C_{cc}-C_{bb}\\-2C_{c\bar{n}}+2C_{b\bar{n}}\end{pmatrix}&
\begin{pmatrix}\frac{10}{3}C_{cc}+\frac{10}{3}C_{bb}-4C_{cb}\\-\frac23C_{c\bar{n}}-\frac23C_{b\bar{n}}\end{pmatrix}&
-\frac{2}{3\sqrt{3}}\begin{pmatrix}C_{nn}-C_{bb}+\\7C_{c\bar{n}}-7C_{b\bar{n}}\end{pmatrix}&
2\sqrt{10}(C_{c\bar{n}}+C_{b\bar{n}}) \\

-\frac{2\sqrt{2}}{9}\begin{pmatrix}C_{cc}+C_{bb}+2C_{cb}\end{pmatrix}&
-\frac{2}{3\sqrt{3}}\begin{pmatrix}C_{nn}-C_{bb}+\\7C_{c\bar{n}}-7C_{b\bar{n}}\end{pmatrix}&
\begin{pmatrix}\frac{26}{9}C_{cc}+\frac{26}{9}C_{bb}-\frac{100}{9}C_{cb}\\+\frac{10}{3}C_{c\bar{n}}+\frac{10}{3}C_{b\bar{n}}\end{pmatrix}&
-\frac{2}{3}\sqrt{\frac{10}{3}}(C_{c\bar{n}}-C_{b\bar{n}})\\

\frac{16}{3}\sqrt{\frac53}(C_{c\bar{n}}-C_{b\bar{n}}) &
2\sqrt{10}(C_{c\bar{n}}+C_{b\bar{n}})&
-\frac{2}{3}\sqrt{\frac{10}{3}}(C_{c\bar{n}}-C_{b\bar{n}})&
\begin{pmatrix}\frac83C_{cc}+\frac83C_{bb}+\frac83C_{cb}\\-4C_{c\bar{n}}-4C_{b\bar{n}}\end{pmatrix}
\end{pmatrix}$
\\
$J=1/2$&$\begin{pmatrix}
\begin{pmatrix}\frac{28}{9}C_{cc}+\frac{28}{9}C_{bb}+\frac{112}{9}C_{cb}\\+\frac{16}{3}C_{c\bar{n}}+\frac{16}{3}C_{b\bar{n}}\end{pmatrix}&
\frac{2}{3}\sqrt{\frac23}\begin{pmatrix}C_{cc}-C_{bb}\\+4C_{c\bar{n}}-4C_{b\bar{n}}\end{pmatrix}&
-\frac{2\sqrt{2}}{9}(C_{cc}+C_{bb}+2C_{cb})&
-\frac{4}{3}\sqrt{\frac23}(C_{c\bar{n}}-C_{b\bar{n}})\\

\frac{2}{3}\sqrt{\frac23}\begin{pmatrix}C_{cc}-C_{bb}\\+4C_{c\bar{n}}-4C_{b\bar{n}}\end{pmatrix}&
\begin{pmatrix}\frac{10}{3}C_{cc}+\frac{10}{3}C_{bb}-4C_{cb}\\+\frac43C_{c\bar{n}}+\frac43C_{b\bar{n}}\end{pmatrix}&
-\frac{2}{3\sqrt{3}}\begin{pmatrix}C_{cc}-C_{bb}-14\\C_{c\bar{n}}+14C_{b\bar{n}}\end{pmatrix}&
-4(C_{c\bar{n}}+C_{b\bar{n}}) \\

-\frac{2\sqrt{2}}{9}(C_{cc}+C_{bb}+2C_{cb})&
-\frac{2}{3\sqrt{3}}\begin{pmatrix}C_{cc}-C_{bb}-14\\C_{c\bar{n}}+14C_{b\bar{n}}\end{pmatrix}&
\begin{pmatrix}\frac{26}{9}C_{cc}+\frac{26}{9}C_{bb}-\frac{100}{9}C_{cb}\\-\frac{20}{3}C_{c\bar{n}}-\frac{20}{3}C_{b\bar{n}}\end{pmatrix}&
\frac{28}{3\sqrt3}(C_{c\bar{n}}-C_{b\bar{n}})\\

-\frac{4}{3}\sqrt{\frac23}(C_{c\bar{n}}-C_{b\bar{n}}) &
-4(C_{c\bar{n}}+C_{b\bar{n}})&
\frac{28}{3\sqrt3}(C_{c\bar{n}}-C_{b\bar{n}})&
\frac83(C_{cc}+C_{bb}-2C_{cb})
\end{pmatrix}$\\
\bottomrule[0.5pt]
\midrule[1.5pt]
\end{tabular}
\end{lrbox}\scalebox{0.93}{\usebox{\tablebox}}
\end{table*}


\end{document}